\documentclass[aps,prb,groupedaddress,twocolumn,showpacs,superscriptaddress,amssymb,amsmath]{revtex4-1}

\usepackage{algorithm}
\usepackage{algpseudocode}

\usepackage{amsmath}
\usepackage{physics}
\usepackage{verbatim}
\usepackage{graphicx}
\usepackage{qting}
\usepackage{amssymb}
\usepackage{amsfonts}
\usepackage{color}
\usepackage{bm}        
\usepackage{comment}
\usepackage{placeins}
\usepackage[mathscr]{eucal}
\usepackage[colorlinks, linkcolor=red,citecolor=blue,urlcolor=blue]{hyperref}
\usepackage[normalem]{ulem}
\usepackage[all]{hypcap}
\usepackage{multirow}
\usepackage[dvipsnames,usenames,table]{xcolor}
\usepackage{dcolumn}
\usepackage[capitalise]{cleveref}
\usepackage{hyperref}
\usepackage{bm}
\usepackage{epsf}
\usepackage{subfigure}
\usepackage{amsfonts}
\usepackage{epstopdf}%

\usepackage{environ}
\usepackage{pifont}
\usepackage{xcolor}
\newcommand{\cmark}{\textcolor{green!80!black}{\ding{51}}}
\newcommand{\xmark}{\textcolor{red}{\ding{55}}}
\makeatletter
\newsavebox{\measure@tikzpicture}
\NewEnviron{scaletikzpicturetowidth}[1]{%
  \def\tikz@width{#1}%
  \begin{lrbox}{\measure@tikzpicture}%
  \BODY
  \end{lrbox}%
  \pgfmathparse{#1/\wd\measure@tikzpicture}%
  \BODY
}

\newcommand{\be}{\begin{equation}}
\newcommand{\ee}{\end{equation}}
\newcommand{\bea}{\begin{eqnarray}}
\newcommand{\eea}{\end{eqnarray}}

\definecolor{mblue}{rgb}{0.0,0.45,0.74}
\usepackage{hyperref}
\usepackage{braket}
\usepackage{amsfonts}
\usepackage{tikz}
\usepackage{pgfplots}
\pgfplotsset{compat=1.15}
\usepackage{mathrsfs}
\usetikzlibrary{arrows}

\allowdisplaybreaks
\begin{document}
\title{Unified Framework for Open Quantum Dynamics with Memory}
\author{Felix Ivander}
\affiliation{Quantum Science and Engineering, Harvard University, Cambridge, MA, USA}
\author{Lachlan P. Lindoy}
\affiliation{National Physical Laboratory, Teddington, TW11 0LW, United Kingdom}
\author{Joonho Lee}
\email{joonholee@g.harvard.edu}
\affiliation{Department of Chemistry and Chemical Biology, Harvard University, Cambridge, MA, USA}
\affiliation{Google Quantum AI, Venice, CA, USA}
\begin{abstract}
Studies of the dynamics of a quantum system coupled to baths are typically performed by utilizing the Nakajima-Zwanzig memory kernel ($\bm{\mathcal{K}}$) or the influence functions ($\mathbf{{I}}$), especially when the dynamics exhibit memory effects (i.e., non-Markovian). Despite their significance, the formal connection between the memory kernel and the influence functions has not been explicitly made. We reveal their relation by inspecting the system propagator for a broad class of problems where an $N$-level system is linearly coupled to Gaussian baths (bosonic, fermionic, and spin.) 
With this connection, we also show how approximate path integral methods can be understood in terms of approximate memory kernels.
For a certain class of open quantum system problems, we devised a non-perturbative, diagrammatic approach to construct $\bm{\mathcal{K}}$ from $\mathbf{{I}}$ for (driven) systems interacting with Gaussian baths without the use of any projection-free dynamics inputs required by standard approaches.
Lastly, we demonstrate a Hamiltonian learning procedure to extract the bath spectral density from a set of reduced system trajectories obtained experimentally or by numerically exact methods, opening new avenues in quantum sensing and engineering.
The insights we provide in this work will significantly advance the understanding of non-Markovian dynamics, and they will be an important stepping stone for theoretical and experimental developments in this area.
\end{abstract}

\maketitle


\textit{Introduction.} Most existing quantum systems inevitably interact with the surrounding environment, often making a straightforward application of Schr{\"o}dinger's equation impractical.~\cite{Haboubi1992Sep}  
The main challenge in modeling these ``open'' quantum systems is the large Hilbert space dimension because the environment is much larger than the system of interest.
Addressing this challenge is important in many disciplines, including solid state and condensed matter physics,\cite{CMP1,CMP2,CMP3} chemical physics and quantum biology,\cite{QB1,QB2,HEOM2,SpaventaPRA} quantum optics,\cite{andersson_non-exponential_2019,QO1,QO2,QO3} and quantum information science.\cite{NQI1,Bylicka2014,NNE4} 
In this article, we
provide a unified framework for studying non-Markovian open quantum systems, which will help to facilitate a better understanding of open quantum dynamics and the development of numerical methods.

Various numerically exact methods have been developed to describe {\it non-Markovian} open quantum dynamics. Two of the most commonly used approaches are (1) the Feynman-Vernon influence functional path integral (INFPI)~\cite{FeynmanVernon} based techniques, including the quasiadiabatic path-integral method of Makri and Makarov and its variants \cite{quapi2,quapi3,smatpi1,smatpi2,Pathsum,Segal2007,INFPI1,INFPI2,MPI,QCPI}, hierarchical equations of motion (HEOM) methods,~\cite{HEOM1,HEOM2,FPHEOM} and time-evolving matrix product operator and related process tensor-based approaches
,~\cite{TEMPO2018,TEBD2,ITEBD_PT,TEMPO2,jorgensen2019,Jorgensen2020Nov} 
and (2) the Nakajima-Zwanzig generalized quantum master equation (GQME) techniques. \cite{BPBook,nakajima,zwanzig,Mulvihill2021Sep} 
The INFPI formulation employs the influence functional ($\mathcal I$) that encodes the time-nonlocal influence of the baths on the system. In the GQME formalism, the analogous object to $\mathcal I$ is the memory kernel ($\bm{\mathcal{K}}$), which describes the entire complexity of the bath influence on the reduced system dynamics. It is natural to intuit that $\mathcal I$ and $\bm{\mathcal{K}}$ are closely connected and are presumably identical in their information content. Despite this, to the best of our knowledge, \textit{analytic} and \textit{explicit} relationships between the two have yet to be shown. 

\begin{figure*}[hbt!]
    \centering
    \includegraphics[width=1.7\columnwidth]{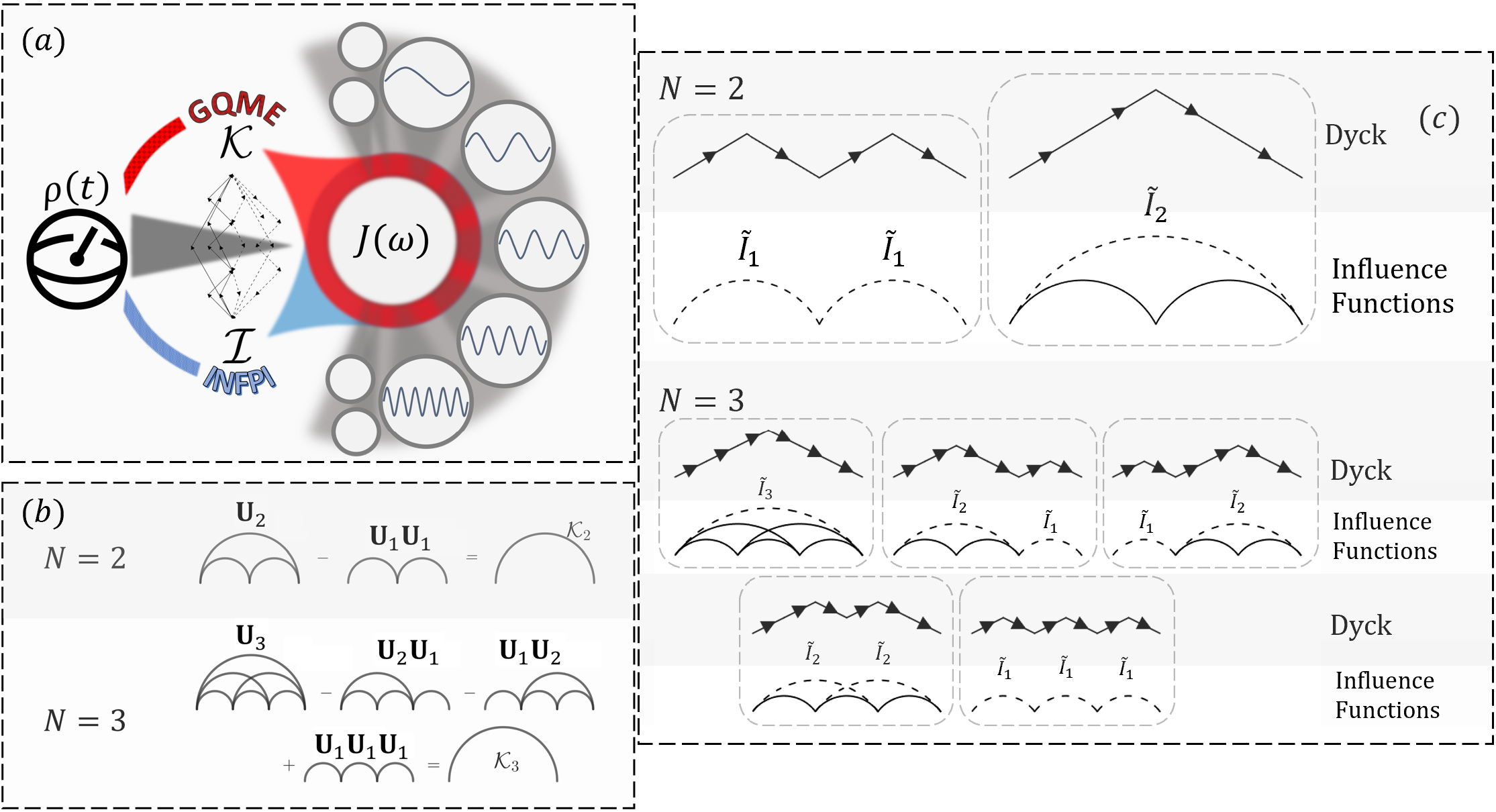}
    \caption{\textbf{Unification of open quantum dynamics framework for Class 1.} Panel (a): An open quantum system, where the environment is characterized by the spectral density $J(\omega)$, can be described with the Generalized Quantum Master Equation (GQME) and the Influence Functional Path Integral (INFPI). The former distills environmental correlations through the memory kernels $\mathcal{K}$ while the latter through the influence functionals $\mathcal{I}$. In this work, we show both are related through Dyck Paths, and that furthermore we can use the Dyck construction for extracting $J(\omega)$ by simply knowing how the quantum system evolves. Panel (b): {Cumulant expansion of memory kernel.} Examples through Eq. (\ref{mEMTTIM}) for $N=2$ and $N=3$. Solid arcs of diameter $k$ filled with all possible arcs of diameters smaller than $k$ denote propagator $\mathbf{U}_k$. Panel (c): Dyck path diagrams. Examples for $N=2$ and $N=3$ and their corresponding influence function diagrams, which composes $\bm{\mathcal{K}}_2$ and $\bm{\mathcal{K}}_3$ respectively. Solid lines denote influence functions ${\mathbf I}$ and dashed line denote $\tilde{\mathbf I}$.}
    \label{D1}
\end{figure*}

There have been several works that loosely connect these two frameworks. For instance, there is a body of work on numerically computing $\bm{\mathcal{K}}$ with projection-free inputs using short-time system trajectories based on INFPI or other exact quantum dynamics methods.~\cite{shi2003,shi2004,CaoTTM,Kidon2018Sep,golosov1999efficient} The obtained $\bm{\mathcal{K}}$ is then used to propagate system dynamics to longer times. 
Another line of work worth noting is 
the real-time path integral Monte Carlo algorithms
for evaluating memory kernels exactly.\cite{Cohen2011Aug}
These works took advantage of the real-time path integral approaches used to evaluate $\mathcal I$ \cite{Muhlbacher2008May} to evaluate necessary matrix elements in computing the exact memory kernel.
Nonetheless, they did not present any direct analytical relationship between
the memory kernel and $\mathcal I$.
In this Article, we present a unifying description of these non-Markovian quantum dynamics frameworks. In particular, we establish explicit {\it analytic correspondence} between $\mathcal I$ and $\bm{\mathcal{K}}$. We present a visual schematic describing the main idea of our work in \cref{D1} panel (a). Readers interested in the relationship between our work and existing numerical tools are referred to \cref{subsec:UinI}. 

\textit{Overview of our contribution.}
We consider a broad range of system-bath Hamiltonians in which the bath is Gaussian and the system-bath Hamiltonian is bilinear.
The total Hamiltonian is $\hat{H} = \hat{H}_S + \sum_j(\hat{H}_{B,j}+ \sum_\alpha\hat{H}_{I,j,\alpha})$, with subscripts $j$ and $\alpha$ specifying the $j$-th bath and the $\alpha$-th interaction, respectively. While we do not limit the form of $\hat{H}_S$ in our discussion, we consider a quadratic (i.e., Gaussian) Hamiltonian for the baths,
$\hat{H}_{B,j}=\sum_k \omega_{k,j} \hat{a}_{k,j}^\dagger\hat{a}_{k,j}$, where $\hat{a}_{k,j}$ can be fermionic or bosonic (it is also possible to treat baths consisting of noninteracting spins in a certain limit, see \ref{sec:class1}), and the bilinear interaction Hamiltonian, $\hat{H}_{I,j,\alpha}=\hat{S}_{j,\alpha}\otimes\hat{B}_{j,\alpha}$ with $\hat{S}_{j,\alpha}$ and $\hat{B}_{j,\alpha}$ being the system and bath operators, respectively. We also assume that the initial density matrix is separable between the system and each bath.
There are four classes of problems that one may commonly encounter under this setup:
\q{classes}{
\begin{enumerate}
    \item \textit{Class 1:} With only single $\alpha$ for all baths $j$ (such cases are henceforth indicated by dropping the subscript $\alpha$), $\{\hat{S}_j\}$ are all diagonalizable, and furthermore that $\{\hat{S}_j\}$ are all \textit{simultaneously} diagonalizable. That is, all terms in $\{\hat{H}_{I,j}\}$ commute. The spin-boson model, other models in the same universality class, and Frenkel exciton models for photosynthetic systems belong to this class.
    \item \textit{Class 2:} No terms in $\{\hat{S}_{j}\}$ commute but each term in $\{\hat{S}_j\}$ is diagonalizable.
    Generalizing the models in {\it Class 1} to multiple nonadditive baths typically leads to this case. Such systems may arise when considering non-adiabatic dynamics of systems involving strong coupling of electronic degrees of freedom coupled to quantized photonic modes. \cite{PRXQuantum.3.010321}
    \item \textit{Class 3:}
    There are common baths for some $\hat{H}_{I,j,\alpha}$ and $\{\hat{S}_{j,\alpha}\}$ may or may not commute. 
    Examples of such baths arise when considering decoherence in models of coupled qubits. \cite{PhysRevB.82.144423} 
    \item \textit{Class 4:} No terms in $\{\hat{S}_{j}\}$ commute and each term in $\{\hat{S}_j\}$ is not diagonalizable. The Anderson impurity model~\cite{Anderson1961Oct} is representative of this category.
\end{enumerate}}
We show in all three classes that one can relate 
$\mathcal I$ and $\bm{\mathcal{K}}$ analytically.
Furthermore, we show that one can obtain the bath spectral density from the reduced dynamics.
Lastly, for {\it Class 1}, we show that a simple diagrammatic structure in the relationship between $\mathcal I$ and $\bm{\mathcal{K}}$ can be found, which allows for efficient construction of $\bm{\mathcal{K}}$ without approximations.
We provide more details of {\it Class 1} in the main text, and additional details for other classes are available in the Appendix.  Further, for {\it Class 1} models, we extend this analysis to consider driven systems, extending the analysis beyond the time-translationally invariant memory kernels observed for time-independent Hamiltonians.

\textit{Path Integral Formulation.}
The time evolution of the full system is given by, $\rho_\text{tot}(t)=e^{-i\hat{H}t}\rho_\text{tot}(0)e^{i\hat{H}t}$. 
We discretize time and employ a Trotterized propagator,
\bea
e^{-i\hat{H}\Delta t} = e^{-i\hat{H}_S \Delta t/2} e^{-i\hat{H}_\text{env} \Delta t} e^{-i\hat{H}_S \Delta t/2} + O(\Delta t^3),
\eea
where $\hat{H}_\text{env}=\hat{H}-\hat{H}_S$. 
The initial total density matrix is assumed to be factorized into 
$\rho_\text{tot}(0) = \rho(0) \otimes (Z_j^{-1}\exp\left[-\beta_j \hat{H}_{B,j}\right])^{\otimes j}$ at inverse temperature $\beta_j$ where $Z_j = \Tr{\exp\left[-\beta \hat{H}_{B,j}\right]}$.
Then, one can show that the dynamics of the reduced system density matrix, $\rho(N\Delta t) = \rho_{N} = \text{Tr}_B\left[\rho_\text{tot}(N\Delta t)\right]$ (partial trace over all baths' degree of freedom), follows 
\begin{align}\nonumber
\langle x^+_{2N}|\rho_{N}|x^-_{2N}\rangle=&
\sum_{x^\pm_0\cdots x^\pm_{2N-1}}
G_{x^\pm_{0} x^\pm_{1}}
G_{x^\pm_{1} x^\pm_{2}}
\dots G_{x^\pm_{2N-1} x^\pm_{2N}}\\
&\times \langle x^+_0|\rho_{0}|x^-_0\rangle\prod_\alpha \mathcal I_j(x_1^\pm,x_3^\pm,\cdots, x_{2N-1}^\pm),
\label{eq:infpi}
\end{align}
where $G_{x^\pm_m x^\pm_{m+1}}=\langle x^+_m|e^{-\frac{i\hat{H}_{s} \Delta t}{2}}| x^+_{m+1}\rangle\langle x^-_{m+1}|e^{\frac{i\hat{H}_{s} \Delta t}{2}}| x^-_{m}\rangle$. 

Restricting ourselves to problems in \textit{Class 1} (details for other \textit{Classes} are available in the Appendix), 
 we consider $\hat{H}_{I}=\hat{S}\otimes\hat{B}$ where $\hat{S}$ is a system operator that is diagonal in the computational basis and $\hat{B}=\sum_k \lambda_k(\hat{a}^\dagger_k+\hat{a}_k)$ is a bath operator that is linear in the bath creation and annihilation operators (with the subscript $\alpha$ and $j$ dropped for clarity.) 
The discussion below can be applied to cases with multiple commuting $\hat{S}\otimes\hat{B}$ since $\mathcal{I}$ take simple product form, see \ref{sec:pi}.
We can show that the influence functional, $\mathcal I$, is pairwise separable,
\begin{align}\nonumber
\mathcal I(x_1^\pm,x_3^\pm,\cdots, x_{2N-1}^\pm)
=&\prod_{n=1}^{N}I_{0,x_{2n-1}^\pm}
\prod_{n=1}^{N-1}I_{1,x_{2n-1}^\pm x_{2n+1}^\pm}
\nonumber\\ \nonumber
&
\times \prod_{n=2}^{N-1}I_{2,x_{2n-3}^\pm x_{2n+1}^\pm}\\
&\cdots \times I_{N-1,x_{1}^\pm x_{2N-1}^\pm}
\label{eq:IF}
\end{align}
where the {\it influence functions} $\mathbf{{I}}_k$ are defined in \ref{sec:pi}, and are related to the bath spectral density, $J(\omega)=\pi\sum_k\lambda^2_k\delta(\omega-\omega_k)$. 
For later use, we note that
\cref{eq:infpi} can be simplified into
\begin{equation}
\langle x^+_{2N}|\rho_{N}|x^-_{2N}\rangle
=
\sum_{x^\pm_0}
(\mathbf U_N)_{x^\pm_{2N}x^\pm_0}
\langle x^+_0|\rho_{0}|x^-_0\rangle,
\label{eq:UN}
\end{equation}
where $\mathbf U_N$ is the system propagator
from $t=0$ to $t=N\Delta t$.
It is then straightforward to express $\mathbf U_N$ in terms of $\{\mathbf{I}_k\}$.~\cite{golosov1999efficient,smatpi1,smatpi2,Pathsum,tree-smatpi}

\textit{The Nakajima-Zwanzig Equation.} The Nakajima-Zwanzig equation is a time-non-local formulation of the formally exact GQME. Assuming the time-independence of $\hat{H}_S$, the discretized homogeneous Nakajima-Zwanzig equation takes the form
\bea
\rho_{N}={\mathbf{L}}\rho_{N-1}+\Delta t^2\sum_{m=1}^{N} \bm{\mathcal{K}}_{N-m}\rho_{m-1}, \label{GQME}
\eea
where 
$\mathbf L \equiv (\mathbf{1}-\frac{i}{\hbar}\mathcal{L}_S\Delta t)$ with
$\mathcal{L}_S \bullet\equiv[\hat{H}_S,\bullet]$ being the bare system Liouvillian and $\bm{\mathcal{K}}_{n}$ is the discrete-time memory kernel at timestep $n$. To relate $\bm{\mathcal{K}}_N$ to $\{\mathbf{{I}}_k\}$, we inspect the reduced dynamics evolution operator $\mathbf U_N$ as defined in \cref{eq:UN},
\begin{equation}
\mathbf{U}_N=\mathbf{L}\mathbf{U}_{N-1}+\Delta t^2\sum_{m=1}^{N}\bm{\mathcal{K}}_{N-m}\mathbf{U}_{m-1}. \label{mEMTTIM}
\end{equation}
With this relation, one can obtain $\bm{\mathcal{K}}_N$ from the reduced propagators $\{\mathbf U_k\}$. We observe setting $N=1$ yields 
$\bm{\mathcal{K}}_{0}=\frac{1}{\Delta t^2}(\mathbf{U}_1-\mathbf{L})$, since $\mathbf{U}_0$ is the identity.
The memory kernel, $\bm{\mathcal{K}}_{0}$, accounts for
the deviation of the system dynamics from its pure dynamics (decoupled from the bath) within a time step.
From setting $N=2$, we get
$\bm{\mathcal{K}}_{1}=\frac{1}{\Delta t^2}(\mathbf{U}_2-\mathbf{U}_1\mathbf{U}_1)$. 
This intuitively shows that $\bm{\mathcal{K}}_{1}$ captures the effect of the bath that cannot be captured within $\bm{\mathcal{K}}_{0}$.
Similarly, for $N=3$,
$
\bm{\mathcal{K}}_{2}=\frac{1}{\Delta t^2}(\mathbf{U}_3-\mathbf{U}_2\mathbf{U}_1-\mathbf{U}_1\mathbf{U}_2+\mathbf{U}_1\mathbf{U}_1\mathbf{U}_1).
$
This set of equations 
is similar to 
cumulant expansions, widely used in  
many-body physics and electronic structure theory.\cite{Mahan,Kutzelnigg1999Feb}
Instead of dealing with higher-order $N$-body expectation values,
we deal with higher-order $N$-time memory kernel in this context.
The $N$-time memory kernel $\bm{\mathcal{K}}_N$ is the $N$-th order cumulant in the cumulant expansion of the system operator.
Unsurprisingly, these recursive relations lead to
diagrammatic expansions commonly found in cumulant expansions\cite{Mahan}
as shown in \cref{D1} panel (b).

{\it Main results.} Using this cumulant generation of $\bm{\mathcal{K}}_N$ and by expressing $\{\mathbf U_k\}$ in terms of $\{\mathbf{{I}}_k\}$, we obtain a direct relationship between $\bm{\mathcal{K}}_N$ and $\{\mathbf{{I}}_k\}_{k=0}^{k=N}$. Specifically, we have
\begin{align}
{\mathcal{K}}_{0,ik}&=\frac{1}{\Delta t^2}\Big[\sum_jG_{ij}I_{0,j}G_{jk}-L_{ik}\Big] \ \ \label{eq:K0}\\
{\mathcal{K}}_{1,im}&=
\frac{1}{\Delta t^2}\sum_{jk}G_{ij}I_{0,j}F_{jk}\tilde{I}_{1,jk}I_{0,k}G_{km}\label{eq:K1}\\
{\mathcal{K}}_{2,ip}&=\frac{1}{\Delta t^2}\sum_{jkn}
G_{ij}F_{jk}F_{kn}(\tilde{I}_{2,jn}I_{1,jk}I_{1,kn}\nonumber\\
&+\tilde{I}_{1,jk}\tilde{I}_{1,kn})I_{0,j}I_{0,k}I_{0,n}G_{np}\label{eq:K2}\\
{\mathcal{K}}_{3,il}&=\frac{1}{\Delta t^2}\sum_{jknp}
G_{ij}F_{jk}F_{kn}F_{np}I_{0,j}I_{0,k}I_{0,n}I_{0,p}G_{pl}\nonumber\\
&\Big\{\tilde{I}_{3,jp}I_{2,jn}I_{2,kp}I_{1,jk}I_{1,kn}I_{1,np}\nonumber\\
&+
I_{1,kn}
\big(\tilde{I}_{2,jn}\tilde{I}_{2,kp}I_{1,jk}I_{1,np}
+\tilde{I}_{2,kp}\tilde{I}_{1,jk}I_{1,np}\nonumber\\
&\ \ \ +\tilde{I}_{2,jn}\tilde{I}_{1,np}I_{1,jk}\big)+\tilde{I}_{1,jk}\tilde{I}_{1,kn}\tilde{I}_{1,np}\Big\}\label{eq:K3}\\
&\vdots&\nonumber
\end{align} 
where
we define $\mathbf F = \mathbf G \mathbf G$ (bold-face for denoting matrices) 
and $\tilde{I}_{k,ij} = I_{k,ij}-1$. We emphasize that \cref{eq:K0,eq:K1,eq:K2,eq:K3} are exact up to the Trotter discretization error and valid for any coupling strengths in the models considered in this work. By definition, earlier $\mathcal{K}_N$ contains shorter memory effects and will thus appear simpler.

This series of equations is a part of the main result of this work, showing explicitly how $\bm{\mathcal{K}}_N$ is diagrammatically constructed in terms of influence functions from $\mathbf I_0$ to $\mathbf I_{N}$.
This construction can easily show the computational effort of computing $\bm{\mathcal{K}}_N$.
We sum over an additional time index for each time step.
This gives a computational cost that scales exponentially in time, $\mathcal O(N_\text{dim}^{2N})$
where $N_\text{dim}$ is the dimension of the system Hilbert space. 
In \ref{sec:Dyck}, we present further details on the general algorithm for calculating higher-order memory kernels, exploiting a non-trivial diagrammatic structure to express them in terms of $\mathbf{I}$ and $\tilde{\mathbf{I}}$.

It can be inferred from \cref{eq:K1,eq:K2,eq:K3} that each term in $\bm{\mathcal{K}}_N$ is represented uniquely by each Dyck path \cite{brualdi2010introductory,catalanbook,catalanoies} of order $N$. Hence, one can construct $\bm{\mathcal{K}}_N$ by generating the respective set of Dyck paths and associating each path with a tensor contraction of influence functions. This is illustrated in Fig.~\ref{D1} panel (c) and further detailed in \ref{sec:Dyck}.
This observation reveals some new properties of $\bm{\mathcal{K}}_N$. First, the number of terms in $\bm{\mathcal{K}}_{N}$ is given by the $N$-th Catalan's number\cite{catalanbook,catalanoies} $C_N=\frac{1}{N+1}{{2N}\choose{N}}$ (i.e., $\bm{\mathcal{K}}_{4}$ has 14 such terms, $\bm{\mathcal{K}}_{5}$ has 42, then 132, 429, 1430, 4862, 16796, 58786, $\dots$). We note that Catalan's number appeared in Ref.~\citenum{tree-smatpi} when analyzing an approximate numerical INFPI method.
See \ref{sec:Dyck} for more information.

Scrutinizing the relationship of $\bm{\mathcal{K}}$ and $\mathbf{I}$, presented in \ref{sec:Dyck}, further, we can observe how $\bm{\mathcal{K}}$ decays asymptotically. 
As is well-known, for typical condensed phase systems $I_{k,ij}\to 1$ for $k\to \infty$.\cite{quapi1, quapi2}
Similarly, because $\tilde{I}_{k,ij}\ll 1$ for large $k$, those terms with larger multiplicities contribute less to $\bm{\mathcal{K}}_N$ and decay exponentially to zero as multiplicity grows.
In fact, for condensed phase systems, the decay of $\mathbf{I}_N$ and $\bm{\mathcal{K}}_N$ is often rapid, which motivated the development of approximate INFPI methods\cite{quapi1,quapi2,quapi3,smatpi1,smatpi2} and other approximate GQME methods.\cite{aGQME1,aGQME2,aGQME3,Kelly2016May}

With our new insight, approximate INFPI methods can be viewed through the lens of the corresponding memory kernel content (and vice versa).
As an example, we shall discuss the iterative quasiadiabatic path-integral methods.\cite{quapi1,quapi2,quapi3} In these methods, $I_{k,ij}$ is set to unity beyond a preset truncation length $k_\text{max}$. For simplicity, let us consider $k_\text{max} = 1 $, and hence $I_{k,ij}=1$ and $\tilde{I}_{k,ij}=0$ for $k>k_\text{max}$. We now inspect what this approximation entails for $\bm{\mathcal{K}}_N$. First, no approximation is applied to $\bm{\mathcal{K}}_0$ and $\bm{\mathcal{K}}_1$. Then, in $\bm{\mathcal{K}}_2$ (\cref{eq:K2}), 
\bea
(\tilde{I}_{2,jn}I_{1,jk}I_{1,kn}+\tilde{I}_{1,jk}\tilde{I}_{1,kn})\to\tilde{I}_{1,jk}\tilde{I}_{1,kn}.
\eea 
Similarly, in $\bm{\mathcal{K}}_3$ (\cref{eq:K3}), the only surviving contribution is from $\tilde{I}_{1,jk}\tilde{I}_{1,kn}\tilde{I}_{1,np}$.
We hope such a direct connection between approximate methods will inspire the development of more efficient and accurate methods.

The time-translational structure of the INFPI formulation and its Dyck-diagrammatic structure allow for a recursive deduction of ${\mathbf I}_N$ from $\bm{\mathcal{K}}_N$, which is the inverse map of \cref{eq:K1,eq:K2,eq:K3}. 
We first observe that 
\begin{equation}
\mathbf I_0 = \mathbf{G}^{-1} (\delta t^2 \bm{\mathcal{K}}_0 + \mathbf L)\mathbf{G}^{-1}
\end{equation}
where we obtained $\mathbf I_0$ from $\bm{\mathcal{K}}_0$.
One can then show that
\begin{equation}
I_{1,jk}=1 + \Delta t^2\frac{(\mathbf G^{-1}\bm{\mathcal{K}}_1\mathbf G^{-1})_{jk}}{F_{jk}I_{0,j}I_{0,k}}.\label{eq:I1}
\end{equation}
using $\bm{\mathcal{K}}_1$ and $\mathbf I_0$.
Similarly, inspecting the expression for ${\bm{\mathcal{K}}_2}$ gives us
\begin{align}
I_{2,jn}=1&+\Big[\Delta t^2(\mathbf G^{-1}\bm{\mathcal{K}}_2\mathbf G^{-1})_{jn}\nonumber\\
&\frac{-\sum_{k}
F_{jk}F_{kn}
\tilde{I}_{1,jk}\tilde{I}_{1,kn}I_{0,j}I_{0,k}I_{0,n}\Big]}{\sum_{k}
F_{jk}F_{kn}I_{1,jk}I_{1,kn}
I_{0,j}I_{0,k}I_{0,n}}, \label{eq:I2}
\end{align}
where $\tilde{I}_{1,jk}=I_{1,jk}-1$ as well as $I_{0,i}$ are obtained from the previous two relations. 

\begin{figure*}[hbt!]
    \centering
    \includegraphics[width=2.0\columnwidth]{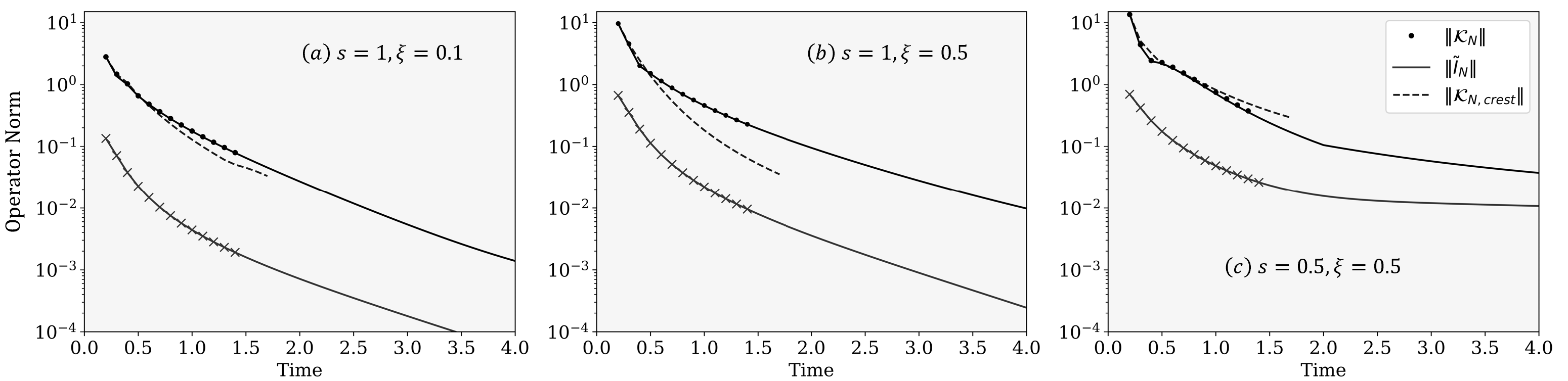}
    \caption{
    \textbf{Verification of the Dyck construction.}
Operator norm of $\tilde{\mathbf I}_N$ (Light) and $\bm{\mathcal{K}}_N$ (Dark) as a function of $N\Delta t$.
    Lines denote $\tilde{\mathbf I}_N$ computed from analytic expressions and $\bm{\mathcal{K}}_N$ from post-processing exact numerical results via the transfer tensor method.\cite{CaoTTM} Circles denote $\bm{\mathcal{K}}_N$ from the Dyck diagrammatic method, and crosses are $\tilde{\mathbf I}_N$ obtained via the inverse map discussed in \cref{eq:I1,eq:I2}. Dashed lines denote the operator norm of the crest term of $\bm{\mathcal{K}}_N$ (the Dyck path diagram with the highest height). Parameters used are: $\Delta=1$ (other parameters are expressed relative to $\Delta$), $\epsilon=0$, $\beta=5$, $\Delta t=0.1$, $\omega_c=7.5$, and $\xi=0.1$ and $s=1$ (panel (a)), 
    }
    \label{Numerics1}
\end{figure*}
 
In \ref{Appendix C F}, we present a general recursive procedure using the Dyck paths and how to obtain the bath spectral density from $I_k$. As a result, we achieve the following mapping from left to right,
\bea
\rho\rightarrow\mathbf{U}\rightarrow\bm{\mathcal{K}}\rightarrow {\mathbf I}\rightarrow J(\omega).
\eea
A remarkable outcome of this analysis is that one can completely characterize the environment (i.e., $J(\omega)$), by inspecting the reduced system dynamics. 
Such a tool is powerful in engineering quantum systems in experiments where we have access to only the reduced system Hamiltonian and reduced system dynamics, but lack information about the environment. Furthermore, this approach provides an alternative to quantum noise spectroscopy.\cite{RevModPhys.89.035002,sung2021multi}
This type of Hamiltonian learning with access only to subsystem observables has been achieved for other simpler Hamiltonians.\cite{Burgarth2011Jan, DiFranco2009May} To our knowledge, our work is the first to show this inverse map for the Hamiltonian considered here. 

Note that the expression \cref{eq:I1} can become ill-defined when $\mathbf{F}$ is diagonal. 
This occurs when $\hat{H}_S$ is diagonal and commutes with $\hat{H}_\text{env}$, constituting a purely dephasing dynamics. In that case, the reduced system dynamics is governed only by the diagonal elements of $\mathbf{{I}}$. Similarly, $\bm{\mathcal{K}}$ is diagonal, as clearly seen in our Dyck path construction. 
As a result, the map $\bm{\mathcal{K}}\leftrightarrow \mathbf{{I}}$ is no longer bijective in that we cannot obtain off-diagonal elements of $\mathbf{{I}}$. Regardless, one can still extract $J(\omega)$ using only the diagonal elements of $\mathbf{{I}}$ via inverse cosine transform. One may worry \cref{eq:I2} could also become ill-conditioned when its denominator vanishes, but $\hat{H}_S$ is not diagonal. If that were the case, the propagator $\mathbf U_2$ would become zero. Therefore, this condition cannot be satisfied in general. Finally, we remark that generalization to extract the $\mathcal{I}_\alpha$ of multiple baths through a single central system is possible and straightforward. See \ref{Appendix C F} for more details.

\textit{Generalization to Driven Systems.} 
While analysis up to this point considered general time-independent systems, in many scenarios, e.g., of biological or engineering relevance, particularly for quantum control applications\cite{MA20211789}, a time-dependent description of the system is necessary.
In such cases, $\bm{\mathcal{K}}$ loses its time-translational properties and should depend on two times. Consequently, \cref{mEMTTIM} cannot be applied. To overcome this, we factorize $\bm{\mathcal{K}}_{N+s,s}$ into time-dependent and time-independent parts. This can be achieved straightforwardly,
as follows: one observes upon the inclusion of time-dependence in $\hat{H}_S$, the terms that are affected in $\bm{\mathcal{K}}_N$, \cref{eq:K0,eq:K1,eq:K2,eq:K3}, are only the bare system propagators $\mathbf{G}$ and $\mathbf{F}$. 
We define the remainder as tensors with $N$ number of indices, $T_{N;x_{s+2},x_{s+4},...,x_{s+2N}}$, which includes all the influence of the bath between $N$ time steps. These tensors only need to be computed once and reused for a later time. Then, one builds the kernels via tensor contraction over two tensors, 
\begin{equation}\label{16}
{\mathcal{K}}_{N+s,s;x_{s+2N+2},x_s}
=\displaystyle
\frac{1}{\Delta t^2}\sum_{\bullet}
P^{N+1+s,s}_{x_s,\bullet,x_{s+2N+2}}
T_{N;\bullet},
\end{equation}
where $\bullet$ denotes indices, $x_{s+2},...,x_{s+2N}$, and the tensor $P^{N+1+s,s}_{x_s,\bullet,x_{2N+s}}$ encapsulates the time-dependence of the system Hamiltonian and is constructed only out of bare system propagators.
The tensor, $T_{N;\bullet}$, then consists only of influence functions, up to $\mathbf{{I}}_{N}$. 
The construction of these tensors is straightforward with $T_{N;\bullet}$ following the Dyck path construction presented for time-independent system dynamics.
\q{PT}{The $T_{N;\bullet}$ tensor appears to be related to the process tensor.~\cite{jorgensen2019,Jorgensen2020Nov} $\mathbf T$ represents $\mathbf K$ upon the contraction with $\mathbf P$, but the process tensor is used to construct $\mathbf U$ when contracted with $\mathbf P$. Subsequently, there is a non-trivial rearrangement of the terms to write $\mathbf K$ in terms of the process tensor. The simple relationship between $\mathbf T$ and $\mathbf K$ in Eq. (\ref{16}) is our unique contribution.}
More detailed analysis and relevant numerical results for open, driven system dynamics are presented in \ref{sec:gaussian}.

\begin{figure*}[hbt!]
    \centering
    \includegraphics[width=2\columnwidth]{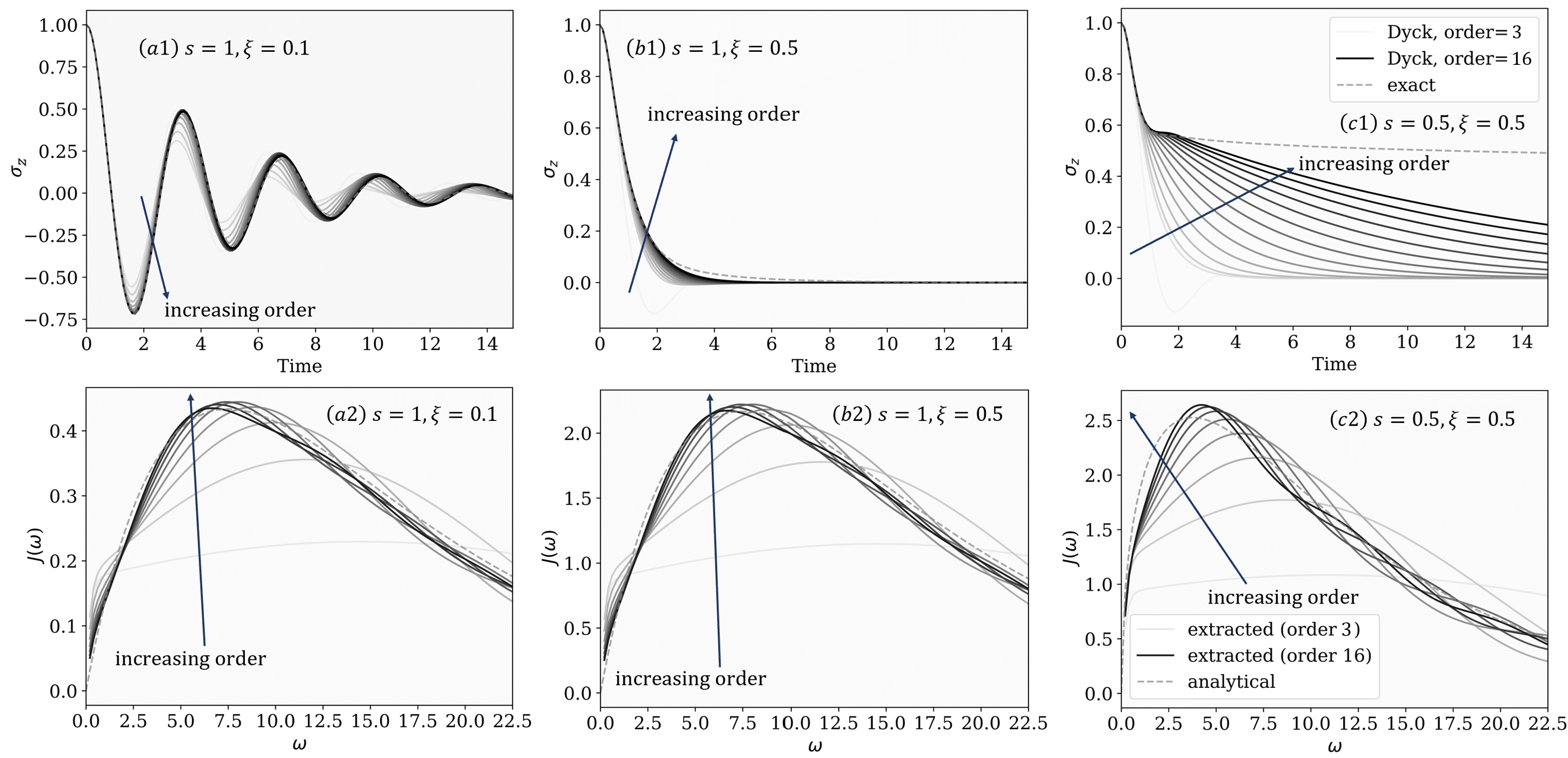}
    \caption{\textbf{Dynamics of spin-boson model with truncated Dyck paths.} Panels (a1), (b1), and (c1): Magnetization ($\langle \sigma_z(t)\rangle$) dynamics predicted using $\bm{\mathcal{K}}$ constructed via Dyck diagrams with increasing truncation orders (from light to darker colors) compared to exact results (see \ref{sec: Appendix F}). Panels (a2), (b2), and (c2): Bath spectral densities extracted through the Dyck diagrammatic method with increasing truncation order (from white to black colors) compared to exact spectral densities (dashed), see \ref{Appendix C F} for more details. These results come from numerically exact trajectories, initiated from linearly independent initial states $\rho_1(0)=\frac{1}{2}(\mathbf{1}+\sigma_z), \ \rho_2(0)=\frac{1}{2}(\mathbf{1}-\sigma_z), \ \rho_3(0)=\frac{1}{2}(\mathbf{1}+\sigma_x), \ \rho_4(0)=\frac{1}{2}(\mathbf{1}+\sigma_x+\sigma_y+\sigma_z)$.
    Parameters used are: $\Delta=1$ (other parameters are expressed relative to $\Delta$), $\epsilon=0$, $\beta=5$, $\Delta t=0.1$ (panels (a1), (b1), and (c1)) or $\Delta t=0.05$ (Panels (a2), (b2), and (c2)), $\omega_c=7.5$, and $\xi=0.1$ and $s=1$ (panels (a1) and (a2)),
    $\xi=0.5$ and $s=1$ (panels (b1) and (b2)), or $\xi=0.5$ and $s=0.5$ (panels (c1) and (c2)).
    }
    \label{Numerics2}
\end{figure*}

{\it Numerical results.} While the discussion above applies to a generic system linearly coupled to a Gaussian bath (or multiple such baths if they couple additively), we discuss the spin-boson model for further illustration. The spin-boson model is an archetypal model for studying open quantum systems.\cite{spinboson1} The model comprises a two-level system coupled linearly to a bath of harmonic oscillators. Hence, it and its generalizations have been used to understand various quantum phenomena: transport, chemical reactions, diode effect, and phase transitions.\cite{Nitzan}

We use $\hat{H}_S=\epsilon\sigma_z+\Delta \sigma_x$, coupled via $\sigma_z$ to a harmonic bath with spectral density ($\omega\geq0$) \cite{spinboson1}
\bea
J(\omega)=\pi\sum_k\lambda^2_k\delta(\omega-\omega_k)=\frac{\xi\pi}{2}\frac{\omega^s}{\omega_c^{s-1}} e^{-\omega/\omega_c},\label{A86C}
\eea
where $J(-\omega)=-J(\omega)$, $\xi$ is the Kondo parameter, and $s$ is the Ohmicity. All reference calculations were performed using the HEOM method.\cite{doi:10.1143/JPSJ.58.101,Tanimura2020Jul, FPHEOM}  Details of the HEOM implementation used here are provided in \ref{sec:HEOM}.

In \cref{Numerics1}, we investigate a series of spin-boson models corresponding to weak and intermediate coupling to an Ohmic environment ($s=1$) as well as strong coupling to a subohmic environment ($s=0.5$). In panels (a) and (b), we observe that the decay of $\tilde{\mathbf{{I}}}_{N}$ is rapid for the Ohmic cases. This translates to a similarly rapid decay for the respective $\bm{\mathcal{K}}_N$, although one can see that both $\tilde{I}_{N}$ and $\bm{\mathcal{K}}_N$ are overall scaled larger in the strong coupling regime. This is to be contrasted with the results for the strongly coupled subohmic environment shown in panel (c). 
The decay of the $\tilde{\mathbf{{I}}}_{N}$ is slow, accompanied by a similarly slow decay of $\bm{\mathcal{K}}_N$. 
Interestingly, the rates by which both $\tilde{\mathbf{{I}}}_{N}$ and $\bm{\mathcal{K}}_N$ decay are similar, which we observe to be exponential.
We also see perfect agreement between $\bm{\mathcal{K}}_N$ constructed from our Dyck diagrammatic method and those obtained by numerically post-processing exact trajectories via the transfer tensor method.~\cite{CaoTTM} Lastly, we construct $\tilde{\mathbf{{I}}}_N$ from $\bm{\mathcal{K}}_N$ up to $N=16$ as exemplified in \cref{eq:I1,eq:I2} and observe perfect agreement between our $\tilde{\mathbf{{I}}}_N$ and those computed from its known analytic formula.

We note that the term with $\tilde{\mathbf{{I}}}_N$ (multiplicity of 1) contributes the most to the memory kernel, $\bm{\mathcal{K}}_N$ for all parameters considered in our work. 
We refer to this term as the ``crest'' term, which corresponds to the Dyck path that goes straight to the top and down straight to the bottom, having the tallest height. 
We see a small difference between the crest term norm and the full memory kernel norm in \cref{Numerics1}, indicating that the memory kernel is dominated by the crest term.
Since the decay of $\tilde{\mathbf{{I}}}_N$ is directly related to the decay of the bath correlation function, one can also make connections between the memory kernel decay and the bath correlation function decay.
Nonetheless, for a stronger system-bath coupling (e.g., \cref{Numerics1} (b)) and for cases
with a long-lived memory (e.g., \cref{Numerics1} (c)), terms other than the crest term contribute non-negligibly, making general analysis of the memory kernel decay challenging.

The cost to numerically compute $\bm{\mathcal{K}}_N$ scales exponentially with $N$. Nevertheless, it is possible to exploit the decay of $\tilde{\mathbf{{I}}}_N$, which is rapid for some environments, e.g., ohmic baths, in turn signifying the decay behavior of $\bm{\mathcal{K}}_N$. This allows truncating the summation in Eq. (\ref{GQME}), enabling dynamical propagation to long times (with linear costs in time) as usually done in small matrix path integral methods \cite{smatpi1,smatpi2} and GQME\cite{CaoTTM} methods. We show in panels (a1) and (b1) of Fig. (\ref{Numerics2}) that this procedure applied to a problem with a rapidly decaying $\bm{\mathcal{K}}_N$ quickly converges to the exact value with a reasonably low-order.
On the other hand, for environments with slowly decaying  $\tilde{\mathbf{{I}}}_N$, the truncation scheme struggles to work effectively. For a strongly coupled subohmic environment, as shown in Fig. (\ref{Numerics2}) panel (c1), one would need truncation orders beyond the current computational capabilities of our implementation (about 16) to converge to the exact value. 
Nonetheless, this illustrates that our direct construction of $\bm{\mathcal{K}}_N$ can recover exact dynamics if sufficiently high-order is used. Furthermore, the construction is non-perturbative and can be applied to strong coupling problems.
\q{QPTTEMPO}{
We note that describing quantum phase transitions at $T=0$ would require capturing the algebraic decay in $\mathbf I_N$.~\cite{TEMPO2018}  Our analysis can, in principle, capture such a slow decay as our approach is exact but will require further optimization in the underlying numerical algorithms for practical applications.}

Finally, in Fig. (\ref{Numerics2}) panels (a2), (b2), and (c2), we show the extraction of spectral densities $J(\omega)$ for three distinct environments. The extracted $J(\omega)$ converges to the analytical value as we obtain the influence functions to higher orders. 
This shows that we can indeed invert the reduced system dynamics to obtain $J(\omega)$ given the knowledge of the system Hamiltonian, which ultimately characterizes the entire system-bath Hamiltonian. 
Nonetheless, the accuracy of the resulting $J(\omega)$ depends on the highest order of $\mathbf{{I}}_k$ we can numerically extract. The cost of extracting $\mathbf{{I}}_k$ scales exponentially in $k$ without approximations, so there is naturally a limit to the precision of $J(\omega)$ in practice.
Furthermore, we show how this procedure can extract highly structured spectral densities as well in \ref{sec: Appendix F} and \cref{fig:structured}. New opportunities await in using approximately inverted $\mathbf{{I}}_k$ and quantifying the error in the resulting $J(\omega)$.

\textit{Conclusions.} In this Article, we provide analytical analysis along with numerical results that show complete equivalence between the memory kernel ($\bm{\mathcal{K}}$) in the GQME formalism and the influence function ($\mathbf{{I}}$) used in INFPI. Our analysis applies to a broad class of general (driven) systems interacting bilinearly with Gaussian baths.
Furthermore, we showed that one can extract the bath spectral density from the reduced system dynamics with the knowledge of the reduced system Hamiltonian $\hat{H}_S$. 
\q{cMPS}{We believe that this unified framework for studying non-Markovian dynamics will facilitate the development of new analytical and numerical methods that combine the strengths of both GQME and INFPI. For example, deep connections between the present work and recent matrix product state (MPS)-based approaches invite ideas that would efficiently extract the environmental spectral density from reduced system dynamics.~\cite{TEMPO2018,ITEBD_PT,TEMPO2,jorgensen2019,Jorgensen2020Nov}}

\textit{Acknowledgements.} F.I. and J.L. were supported by Harvard University's startup funds. L.P.L acknowledges the support of the Engineering and Physical Sciences
Research Council [grant EP/Y005090/1]. We thank Nathan Ng, David Reichman, Dvira Segal, and Jonathan Keeling for stimulating discussions, Tom O'Brien for discussions on Hamiltonian learning, and Hieu Dinh for providing a code to generate the Dyck path. Computations were carried out partly on the FASRC cluster supported by the FAS Division of Science Research Computing Group at Harvard University. This work also used the Delta system at the National Center for Supercomputing Applications through allocation CHE230078 from the Advanced Cyberinfrastructure Coordination Ecosystem: Services \& Support (ACCESS) program, which is supported by National Science Foundation grants \#2138259, \#2138286, \#2138307, \#2137603, and \#2138296. 
\bibliography{refs}
\begin{widetext}
\renewcommand{\theequation}{A\arabic{equation}}
\renewcommand{\thesection}{Appendix \Alph{section}}
\setcounter{equation}{0}  
\setcounter{section}{0}
\section{Path Integral Formulation for Gaussian Baths}\label{sec:pi}
We present a more detailed formulation of the influence-functional-based path-integral (INFPI) approaches.
\subsection{General formulation}
We consider the general Hamiltonian of a quantum system interacting with external degrees of freedom,
\bea
\hat{H}=\hat{H}_S+\sum_j\left(\hat{H}_{B,j}+\sum_\alpha\hat{H}_{I,j,\alpha}\right),\label{genham}
\eea
where $\hat{H}_S$ is the system Hamiltonian, $\hat{H}_{B,j}=\sum_k \omega_{k,j} \hat{a}_{k,j}^\dagger\hat{a}_{k,j}$, and
$\hat{H}_{I,j,\alpha}=\hat{S}_{j,\alpha}\otimes\hat{B}_{j,\alpha}$ where $\hat{S}_{j,\alpha}$ is a system operator and $\hat{B}_{j,\alpha}$ 
is a bath operator for the $\alpha$-th interaction term.
The time evolution of the density matrix of the full system, $\rho_\text{tot}(t)$, 
follows 
\bea
\rho_\text{tot}(t)=e^{-i\hat{H}t}\rho_\text{tot}(0)e^{i\hat{H}t}.
\eea
In INFPI, the real-time propagator is evaluated by path-integral.

With the quasiadiabatic approximation, one may split the total Hamiltonian as (here $\hat{H}_\text{env}=\hat{H}-\hat{H}_S=\sum_j\hat{H}_{B,j}+\sum_\alpha\hat{H}_{I,j,\alpha}$)
\bea
e^{-i\hat{H}\Delta t}\simeq e^{-i\hat{H}_S \Delta t/2} e^{-i\hat{H}_\text{env} \Delta t} e^{-i\hat{H}_S \Delta t/2} + O(\Delta t^3).
\eea
With this identity, we express the trotterized evolution as (a summation over the paths implied)
\bea
\langle x_{2N}^+|\rho_\text{tot}(t)|x_{2N}^-\rangle&=&\langle x_{2N}^+| e^{-i\hat{H}_{S} \Delta t/2} e^{-i\hat{H}_\text{env} \Delta t} e^{-i\hat{H}_{S} \Delta t/2}|x^+_{2N-2}\rangle\nonumber\\
&&\cdots \langle x_0^+|\rho_\text{tot}(0)|x^-_0\rangle \cdots\langle x_{2N-2}^-|e^{i\hat{H}_{S} \Delta t/2} e^{i\hat{H}_\text{env} \Delta t} e^{i\hat{H}_{S} \Delta t/2}|x_{2N}^-\rangle.
\eea
We further insert resolution-of-the-identities between every $e^{\pm i\hat{H}_{S} \Delta t/2}$ and $e^{i\hat{H}_\text{env} \Delta t}$ to obtain
\bea
\langle x_{2N}^+|\rho_\text{tot}(t)|x_{2N}^-\rangle=\langle x_{2N}^+| e^{-i\hat{H}_{S} \Delta t/2}| x_{2N}^{+'}\rangle\langle x_{2N}^{+'}| e^{-i\hat{H}_\text{env} \Delta t} | x_{2N-2}^{+'}\rangle\langle x_{2N-2}^{+'}| e^{-i\hat{H}_{S} \Delta t/2}|x^+_{2N-2}\rangle\cdots.
\eea
We consider a product-separable initial condition,
\begin{equation}
\rho_\text{tot}(0) = \rho(0) \otimes \Big[\frac{\exp\left[-\beta_j \hat{H}_{B,j}\right]}{\Tr_B\left(\exp\left[-\beta_j \hat{H}_{B,j}\right]\right)}\Big]^{\otimes j}
=
\rho(0)
\otimes
(\rho_{B,eq,j})^{\otimes j}
.
\end{equation}
With this, upon tracing out the bath degrees of freedom, we have
\bea
\langle x_{2N}^+|\rho(t)|x_{2N}^-\rangle=G_{x^\pm_{2N}x^{\pm'}_{2N}}G_{x^{\pm'}_{2N}x^\pm_{2N-1}}\cdots\langle x_0^+|\rho(0)|x^-_0\rangle\Tr_B[
\langle x_{2N}^+| e^{-i\hat{H}_{env} \Delta t}| x_{2N}^{+'}\rangle
\cdots\rho_{B,eq} \cdots 
\langle x_{2N}^{-'}| e^{-i\hat{H}_{env} \Delta t}| x_{2N}^{-}\rangle]
\label{eq:rhot}
\eea
where $\rho(t) = \text{Tr}_B ( \rho_\text{tot}(t))$ and $G_{x^\pm y^\pm}=\langle x^+|e^{-i\hat{H}_{s} \Delta t/2}| y^+\rangle\langle y^-|e^{i\hat{H}_{s} \Delta t/2}| x^-\rangle$. 
The term with the trace over bath degrees of freedom in \cref{eq:rhot}
is the so-called {\it influence functional} (IF).
Up to this point, no assumptions have been made on the form of $\hat{S}_{j,\alpha}$, $\hat{B}_{j,\alpha}$, and the algebra of $\hat{a}_{k,j}$. 

\subsection{Influence Functionals for independent Gaussian Baths}\label{IFG}
The evaluation of the IF for general interacting baths involves complicated numerical procedures.
However, one can make more progress if baths follow Gaussian statistics, are all independent from each other, and correspond to only one interaction coupling term.

In such cases, the calculation of the influence functional for Gaussian baths have been presented elsewhere\cite{FeynmanVernon,PRXQuantum.4.030316,gribben2022using}, but we sketch here briefly for completeness. With the Hamiltonian Eq. (\ref{genham}), the expression for the dynamics of the total density matrix is expressed as, in Liouville space in the interaction picture,
\bea
\frac{d}{dt}\rho_{\text{tot}}(t)=\sum_j\mathcal{L}_{I,j}(t)\rho_{\text{tot}}(t)\label{eq:A8}
\eea
which is solved by
\bea
\rho_{\text{tot}}(t)=\mathcal{T}\prod_je^{\int_0^t\mathcal{L}_{I,j}(\tau)d\tau}\rho_{\text{tot}}(0),
\eea
where the product form is due to the commuting property of integrals under time-ordering\cite{TEMPO2}. 
There, $\mathcal{T}$ denotes the time-ordering operator, and $\mathcal{L}_{I,j}$ is the interaction Hamiltonian (with respect to $H_S$ and $H_{B,j}$) in Liouville space, $\mathcal{L}_{I,j}\bullet=-i[H_{I,j}(t),\bullet]$ (hats omitted). Now, taking the trace over multiple environments, we have
\bea
\rho(t)=\Tr_{B,N}[\dots \Tr_{B,2}[\Tr_{B,1}[\rho_{\text{tot}}(t)]]\dots]=\prod_j\langle\mathcal{T}e^{\int_0^t\mathcal{L}_{I,j}(\tau)d\tau}\rangle_j\rho(0),\label{eq:A10}
\eea
where $\langle\cdot\rangle_j$ denotes taking the average over the environment $j$. The averaged quantity above is precisely the continuous IF. Now, as the trace over each bath separates into an independent term in the product above, we will consider evaluating one term and, in what follows, will drop the subscript $j$. Applying Kubo's generalized cumulant identity\cite{kubo1962generalized} yields
\bea
\langle\mathcal{T}e^{\int_0^t\mathcal{L}_I(\tau)d\tau}\rangle=\sum_{n=0}^\infty\frac{1}{n!}\langle\mathcal{T}{\int_0^td\tau_1\dots\int_0^td\tau_n\mathcal{L}_I(\tau_1)}\dots\mathcal{L}_I(\tau_n)\rangle. \label{KGC}
\eea
Then, an application of Isserlis' Theorem\cite{isserlis1918formula} (a special case of the well-known Wick's Theorem\cite{wick1950evaluation}), which states that if the random variables $X_j$ have zero mean and are normal (i.e., Gaussian statistics) then $\langle X_1\dots X_n\rangle=\sum_{p\in P^2_n}\prod_{\{i,j\}\in p}\langle X_i X_j\rangle$ where the sum run overs all possible pairings of $\{1,\dots,n\}$ and the product over the pairs in $p$, yields
\bea
\langle\mathcal{T}{\int_0^td\tau_1\dots\int_0^td\tau_n\mathcal{L}_I(\tau_1)}\dots\mathcal{L}_I(\tau_n)\rangle&=&\sum_{p\in P^2_n}\mathcal{T}\int_0^td\tau_i\int_0^td\tau_j\prod_{\{i,j\}\in p}\langle \mathcal{L}_I(\tau_i)\mathcal{L}_I(\tau_j)\rangle\nonumber\\
&=&\mathcal{T}\frac{2n!}{(2^nn!)}\Big(\int_0^td\tau_i\int_0^td\tau_j\langle \mathcal{L}_I(\tau_i)\mathcal{L}_I(\tau_j)\rangle \Big)^n.
\eea
Since the odd moments vanish, there we skipped the odd terms (there are $\frac{(2n)!}{(2^nn!)}$ such pair partitions). Furthermore, because of this property, when we reinsert to Eq. (\ref{KGC}) we obtain
\bea
\langle\mathcal{T}e^{\int_0^t\mathcal{L}_I(\tau)d\tau}\rangle&=&\sum_{n=0}^\infty\frac{1}{(2n)!}\mathcal{T}\frac{2n!}{(2^nn!)}\Big(\int_0^td\tau_i\int_0^td\tau_j\langle \mathcal{L}_I(\tau_i)\mathcal{L}_I(\tau_j)\rangle\Big)^n\nonumber\\
&=&\sum_{n=0}^\infty\frac{1}{n!}\Big(\int_0^td\tau_i\int_0^td\tau_j\frac{\langle \mathcal{T}\mathcal{L}_I(\tau_i)\mathcal{L}_I(\tau_j)\rangle}{2}\Big)^n.
\eea
Then, applying Fubini's theorem\cite{fubini1907sugli} as well as recalling that the above is the Taylor expansion of an exponential we finally get
\begin{equation}
\langle\mathcal{T}e^{\int_0^t\mathcal{L}_I(\tau)d\tau}\rangle=e^{\int_0^td\tau \int_0^{\tau}d\tau'\langle \mathcal{T}\mathcal{L}_I(\tau)\mathcal{L}_I(\tau')\rangle}\label{eq:IFc}
\end{equation}
which is the continuous version of the influence functional in the main work. Another way to see this result is to use $\langle e^X\rangle=e^{\langle X\rangle^2/2}$ if $X$ is Gaussian. Note, when moving beyond Gaussian environments, one would need to consider each term beyond the second order in the cumulant expansion of Eq. (\ref{KGC}) as Isserlis' Theorem becomes inapplicable. Thus, the inversion procedure we showed in the main text could be seen as approximating the action of an anharmonic environment by that of a harmonic one.

For a given open-quantum system Hamiltonian, one can write a specific form for \cref{eq:IFc} and evaluate the IF via the discrete path-integral approach.
Once the INFPI expression is discretized, a matrix tensor, $\mathbf I$, known as the influence function, appears.
They are closely related to the bath-bath correlation functions.
We will show the computation of $\mathbf{I}$ from microscopic parameters. We must consider each class through different models as the form of bath-bath correlation functions will slightly differ. 
We summarize the differences between the classes considered in \cref{tab:tabclass}.

\begin{table}[h]\label{tab:tabclass}
\begin{tabular}{|c|c|c|c|}
    \cline{2-4}
    \multicolumn{1}{c|}{} & Diagonalizable $\{\hat{S}_{j,\alpha}\}$ & Simultaneously diagonalizable $\{\hat{S}_{j,\alpha}\}$ & Single $\{\hat{H}_{I,j,\alpha}\}$ for each bath  \\
    \hline
    \textit{Class 1} & \cmark & \cmark & \cmark \\
    \hline
    \textit{Class 2} & \cmark & \xmark & \cmark \\
    \hline
    \textit{Class 3} & \cmark & \xmark & \xmark \\
    \hline
    \textit{Class 4} & \xmark & \xmark & \xmark \\
    \hline
\end{tabular}
\caption{Summary of different {\it Classes} considered in this work.}
\end{table}

\section{The Nakajima-Zwanzig Equation} \label{Appendix B}
The Nakajima-Zwanzig (NZ) equation takes the form\cite{nakajima}
\bea
\dot{\rho}(t)=-\frac{i}{\hbar}\mathcal{L}\rho(t)+\int_0^t d\tau \bm{\mathcal{K}}(t,\tau)\rho(\tau)+\mathcal{I}(t).\label{NZH}
\eea
In this work, the inhomogeneous term $\mathcal{I}(t)$ vanishes due to the factorized initial condition assumption $\rho_\text{tot}(0)=\rho(0)\otimes\rho_{B,j}^{\otimes j}$. The \textit{discretized} form of the homogeneous NZ equation is
(with $t=(N-1)\Delta t$)
\bea
\frac{\rho_{N}-\rho_{N-1}}{\Delta t}=-\frac{i}{\hbar}\mathcal{L}_s\rho_{N-1}+\sum_{m=1}^{N} \bm{\mathcal{K}}_{N,m}\rho_m\Delta t.
\eea
By noting that $\rho_N=U_N\rho_0$ (where $\rho_N = \rho(N\Delta t)$) we get
\begin{equation}
\mathbf{U}_N=\mathbf{L}\mathbf{U}_{N-1}+\Delta t^2\sum_{m=1}^{N}\bm{\mathcal{K}}_{N,m}\mathbf{U}_{m-1}.\label{NZD}
\end{equation}
This will allow obtaining $\bm{\mathcal{K}}$ in terms of $\mathbf I$ by making substitutions to $\mathbf{U}_k$. 

First, we organize equations expressing $\mathbf{U}_N$ in terms of $\bm{\mathcal{K}}_{N,m}$ as follows, for $N=1$,
\bea
\frac{\rho_{1}-\rho_{0}}{\Delta t}&=&-\frac{i}{\hbar}\mathcal{L}_s\rho_{0}+\bm{\mathcal{K}}_{1,1}\rho_0\Delta t
\eea
so
\bea
\mathbf{U}_1=(\mathbf I-\frac{i}{\hbar}\mathcal{L}_s\Delta t)+\bm{\mathcal{K}}_{1,1}\Delta t^2.\label{eq:KKK1}
\eea
For $N=2$,
\bea
\rho_{2}=(\mathbf I-\frac{i}{\hbar}\mathcal{L}_s\Delta t)\rho_{1}+[\bm{\mathcal{K}}_{2,1}\rho_0+\bm{\mathcal{K}}_{2,2}\rho_1] \Delta t^2.\label{eq:KKK2}
\eea
Similarly, for $N=3$,
\bea
\rho_3&=&(\mathbf{I}-\frac{i}{\hbar}\mathcal{L}_s\Delta t)\rho_2+[\bm{\mathcal{K}}_{3,1}\rho_0+\bm{\mathcal{K}}_{3,2}\rho_1+\bm{\mathcal{K}}_{3,3}\rho_2]\Delta t^2.\label{eq:KKK3}
\eea
Note that so far Eq. \ref{eq:KKK1} to \ref{eq:KKK3} are exact and general. 

Inspecting Eq. (\ref{eq:KKK2}), one observes that with the time-translational invariance of the memory kernels (i.e., valid if the system Hamiltonian does not have explicit time-dependence), 
\begin{equation}
    \bm{\mathcal{K}}_{2,1}=\bm{\mathcal{K}}_{3,2}\equiv\bm{\mathcal{K}}_1.\label{u2minu1}
\end{equation} 
We use this and make appropriate substitutions in our previous results.
Then, we obtain
\bea
\bm{\mathcal{K}}_{1}\rho_{0}=\frac{1}{\Delta t^2}\Big[\mathbf{U}_2-\mathbf{U}_1\mathbf{U}_1\Big]\rho_{0},
\eea
or
\begin{equation}
\bm{\mathcal K}_1 = \frac1{\Delta t^2}(
\mathbf{U}_2-\mathbf{U}_1\mathbf{U}_1
)
\label{eq:appendixK1}
\end{equation}
Then, for Eq. (\ref{eq:KKK3}) we reorganize to
\bea
\mathbf{U}_3\rho_0&=&(\mathbf{1}-\frac{i}{\hbar}\mathcal{L}_s\Delta t)\mathbf{U}_2\rho_0+[\bm{\mathcal{K}}_{3,1}+\frac{1}{\Delta t^2}(\mathbf{U}_2-\mathbf{U}_1\mathbf{U}_1)\mathbf{U}_1\nonumber\\
&&+
\frac{1}{\Delta t^2}(\frac{i}{\hbar}\mathcal{L}_s\Delta t-1+\mathbf{U}_1)
\mathbf{U}_2\rho_0\Delta t^2.
\eea
One observes many terms again cancel, leaving
\begin{equation}
\bm{\mathcal{K}}_{2}=\frac{1}{\Delta t^2}\Big[-{\mathbf{U}}_2\mathbf{U}_1+{\mathbf{U}}_1{\mathbf{U}}_1{\mathbf{U}}_1-\mathbf{U}_1{\mathbf{U}}_2+{\mathbf{U}}_3\Big]. 
\label{eq:appendixK2}
\end{equation}

One can proceed in similar fashion for higher order terms. For $N=4$,
\bea
\rho_4&=&(\mathbf{1}-\frac{i}{\hbar}\mathcal{L}_s\Delta t)\rho_3+[\bm{\mathcal{K}}_{4,1}\rho_0+\bm{\mathcal{K}}_{4,2}\rho_1+\bm{\mathcal{K}}_{4,3}\rho_2+\bm{\mathcal{K}}_{4,4}\rho_3]\Delta t^2\nonumber\\
\mathbf{U}_4\rho_0&=&(\mathbf{1}-\frac{i}{\hbar}\mathcal{L}_s\Delta t)\mathbf{U}_3\rho_0+[\bm{\mathcal{K}}_{4,1}+\bm{\mathcal{K}}_{2}\mathbf{U}_1+\frac{1}{\Delta t^2}(\mathbf{U}_2-\mathbf{U}_1\mathbf{U}_1)\mathbf{U}_2\nonumber\\
&&+
\frac{1}{\Delta t^2}(\frac{i}{\hbar}\mathcal{L}_s\Delta t-1-\mathbf{U}_1)
\mathbf{U}_3]\rho_0\Delta t^2.
\eea
This then gives
\begin{equation}
\label{eq:appendixK3}
\bm{\mathcal{K}}_{3}=
\frac1{\Delta t^2}
\Big[{\mathbf{U}}_4+\mathbf{U}_2\mathbf{U}_1\mathbf{U}_1-{\mathbf{U}}_1{\mathbf{U}}_1{\mathbf{U}}_1\mathbf{U}_1+\mathbf{U}_1\mathbf{U}_2\mathbf{U}_1-\mathbf{U}_3\mathbf{U}_1-\mathbf{U}_2\mathbf{U}_2+\mathbf{U}_1\mathbf{U}_1\mathbf{U}_2-\mathbf{U}_1\mathbf{U}_3\Big].
\end{equation}

This series of equations will be the starting point to express $\bm{\mathcal{K}}$ in terms of $\mathbf{I}$ for different models that represent the different {\it Classes} discussed in the main text, in the following Appendix sections. 
We note that expressing $\mathbf K$ in terms of $\mathbf U$ was also noted in the transfer tensor method by Cerrillo and Cao,\cite{CaoTTM} but their approach does not illustrate how to write $\mathbf K$ in terms of $\mathbf I$ in the end.
In the following subsections, we will show that one can write $\mathbf U$ in terms of $\mathbf I$ and therefore write $\mathbf K$ in terms of $\mathbf I$.

\section{Additional details on {\it Class 1}}\label{sec:class1}
To make more progress, we assume that $\hat{S}_j$ is diagonalizable, all $\{\hat{S}_j\}$ are simultaneously diagonalizable. 
We later consider generalizations beyond these assumptions. 
\subsection{Influence functionals for {\it Class 1}}\label{sec:pi1}
Under the assumptions of {\it Class 1},
the system coupling operators share the same eigenbasis. We thus insert the resolution-of-the-identity in this basis between the exponentials. 
The terms $\langle x_{n}| \Big(\prod_\alpha e^{-i\hat{H}_{\text{env},\alpha} \Delta t} \Big) | x_{n}'\rangle$ are evaluated via successive application of the exponential operator to the ket
\bea
e^{-i\hat{H}_{\text{env},\alpha} \Delta t/2}| x_n\rangle=e^{-i{H}_{\text{env},\alpha} (x_n)\Delta t/2}| x_n\rangle,
\eea
and then we use $\langle x_{n}|x_{n}'\rangle=\delta_{x_{n}x_{n}'}$ to obtain
\begin{align}
\langle x_{2N}^+|\rho(t)|x_{2N}^-\rangle&=G_{x^\pm_{2N}x^\pm_{2N-1}}G_{x^\pm_{2N-1}x^\pm_{2N-2}}\cdots\langle x_0^+|\rho(0)|x^-_0\rangle\\
&
\times\prod_\alpha\Tr_{B,\alpha}\left[
 e^{-i\hat{H}_{\text{env},\alpha}(x_{2N-1}^+)\Delta t}\cdots\rho_{B,eq} \cdots e^{i\hat{H}_{\text{env},\alpha}(x^-_{2N-1})\Delta t}\right]
\end{align}
Here, the influence functional is
\begin{equation}
\mathcal I(x_1^\pm,x_3^\pm,\cdots, x_{2N-1}^\pm)
=
\prod_\alpha\Tr_{B,\alpha}\left[
 e^{-i\hat{H}_{\text{env},\alpha}(x_{2N-1}^+)\Delta t}\cdots\rho_{B,eq} \cdots e^{i\hat{H}_{\text{env},\alpha}(x^-_{2N-1})\Delta t}\right].
\end{equation}
We will see how this functional
can be calculated for Gaussian baths.

\subsection{Representative models in \textit{Class 1}}
\subsubsection{Gaussian Bosonic Environment}
Since generalizing the spin to multilevel systems will be straightforward, for clarity, we consider here the spin-boson model (details in the main text), described by 
$\hat{H}_S=\epsilon\sigma_z+\Delta \sigma_x$, coupled via $\sigma_z$ to a harmonic bath with spectral density ($\omega\geq0$) Eq. (\ref{A86C}). Note that although here we consider coupling only to a single environment, as we saw in \ref{sec:pi}, generalizing to multiple additive environments is straightforward.
The spin-boson is paradigmatic in the sense that many other models share its universality class, including the double-quantum-dot model and the low-temperature limit of dissipating molecules.\cite{Nitzan} 
With the spin-boson Hamiltonian, we open the exponent integrand in Eq. (\ref{eq:IFc}),
\bea
\langle \mathcal{L}_I(\tau)\mathcal{L}_I(\tau')\rangle=-(x^+(\tau)-x^-(\tau))(\langle B(\tau)B(\tau') \rangle_Bx^+(\tau')-\langle B(\tau)B(\tau') \rangle_B^*x^-(\tau')),\label{eq:continuousPI}
\eea
where $x^\pm\in\text{pair}\{+1,-1\}$ in this case since the coupling is via $\sigma_z$. For the spin-boson model,
\bea
\langle B(\tau)B(\tau') \rangle_B&=&\Big\langle \sum_j\lambda_j(\hat{b}_j^\dagger e^{i\omega_jt}+\hat{b}_j e^{-i\omega_jt})\sum_k{\lambda_k}(\hat{b}_k^\dagger+\hat{b}_k)\Big\rangle \ \ \ \ \ (\tau-\tau'\to t) \nonumber \\
&=&\sum_{j,k}\lambda_j\lambda_k[\langle \hat{b}_j^\dagger e^{i\omega_jt}\hat{b}_k^\dagger\rangle+\langle \hat{b}_j^\dagger e^{i\omega_jt} \hat{b}_k \rangle +\langle \hat{b}_j e^{-i\omega_jt} \hat{b}_k^\dagger\rangle + \langle \hat{b}_j e^{-i\omega_jt} \hat{b}_k\rangle] \nonumber \\
&=&\sum_{j}\lambda_j^2[\langle \hat{b}_j^\dagger e^{i\omega_jt} \hat{b}_j \rangle +\langle \hat{b}_j e^{-i\omega_jt} \hat{b}_j^\dagger\rangle]\nonumber\\
&=&\sum_{j}\lambda_j^2[e^{i\omega_jt}\langle  \hat{n}(\omega_j) \rangle +e^{-i\omega_jt} \langle \hat{n}(\omega_j)+1\rangle] \label{bcfs}
\eea
We now take the continuous limit,
and with ${n}(\omega)=\frac{1}{e^{\beta \omega}-1}$ so that $2{n}(\omega)+1=\frac{2}{e^{\beta \omega}-1}+1=\frac{e^{\beta \omega}+1}{e^{\beta \omega}-1}=\frac{e^{\beta \omega/2}+e^{-\beta \omega/2}}{e^{\beta \omega/2}-e^{-\beta \omega/2}}$ this expands to 
\bea
\frac{1}{\pi}\int_0^\infty d\omega J(\omega)[\frac{e^{\beta \omega/2}}{e^{\beta \omega/2}-e^{-\beta \omega/2}} e^{-i\omega t}+\frac{e^{-\beta \omega/2}}{e^{\beta \omega/2}-e^{-\beta \omega/2}} e^{i\omega t}]
\eea
If we allow $J(\omega)=-J(-\omega)$, this integral becomes half of an even integrand. Thus
\bea
\langle B(t)B(0) \rangle_B=\frac{1}{2\pi}\int_{-\infty}^\infty d\omega J(\omega)\frac{e^{\beta \omega/2}e^{-i\omega t}}{\sinh{\beta \omega/2}}.
\eea
Now we start discretizing the path (with $t_N=N\Delta t$),
\bea
x^\pm(t)&=&x_0^\pm(t)\Theta(t-t_0)+\sum^N_{k=1}(x^\pm_k-x^\pm_{k-1}\Theta(t-t_k))\label{eq:dicretepath}
\eea
and see that the IFs are pairwise decomposable (two integrals in the exponent become a double summation), yielding
\begin{equation}
I_{k'-k,x_{k}^{\pm} x_{k'}^{\pm}}=\exp{-(x_k^+-x_k^-)(\eta_{kk'}x_{k'}^+-\eta_{kk'}^*x_{k'}^-)},\label{eq:Ikk'}
\end{equation}
where we have also used the time-translational property of the IFs and thus defined them in terms of time differences only. There, the coefficients $\eta_{kk'}$ are obtained by substituting the discretized path Eq. (\ref{eq:dicretepath}) to Eq. (\ref{eq:continuousPI})
\bea
\eta_{kk'}&=&\frac{2}{\pi}\int_{-\infty}^{\infty}d\omega\Big( \frac{J(\omega)}{\omega^2}\frac{\exp{\beta\hbar\omega/2}}{\sinh{\beta\hbar\omega/2}}\sin^2{(\omega\Delta t/2)}e^{-i\omega\Delta t (k-k')}\Big),\label{nkk'}\\
\eta_{kk}&=&\frac{1}{2\pi}\int_{-\infty}^{\infty}d\omega \frac{J(\omega)}{\omega^2}\frac{\exp{\beta\hbar\omega/2}}{\sinh{\beta\hbar\omega/2}}(1-e^{-i\omega\Delta t}).\label{nkk}
\eea
We remark that generalizing to multilevel systems (while still coupled to the bosonic environment), for example, in Frenkel excitonic models, amounts only to letting $x(t)$ take more and different values. 
\subsubsection{Spin-Fermion via Bosonization}
Here we consider a central system, $\hat{H}_S=\epsilon_0 \sigma_z+\Delta \sigma_x$, interacting with a fermionic environment, $\hat{H}_F=\sum_k\omega_k\hat{c}_k^\dagger \hat{c}_k$ through the interaction Hamiltonian $\hat{H}_I=\ketbra{1}{1}\frac{V}{L}\sum_{k,q}c_k^\dagger c_q$. In the low energy description, when the impurity potential does not create a bound state, this model can be recast to a spin-boson Hamiltonian\cite{Demler2013}
\bea
\hat{H}=\big(\epsilon-\int_0^{E_F}\frac{dE}{\pi}\delta(\sqrt{2mE})\big) \hat{\sigma}_z+\Delta \hat{\sigma}_x+\sum_qv_F\abs{q}\hat{b}_q^\dagger \hat{b}_q+V\hat{\sigma}_z\sum_{q>0}\sqrt{\frac{q}{2\pi L}}(\hat{b}^\dagger_q+\hat{b}_q).
\eea
We can thus analyze this model as though it is a spin-boson model. Particularly, the inverse mapping procedure would extract microscopic parameters here, e.g., $V$, since the spectral density is then given by
\bea
J(\omega)=\sum_q 4V^2\frac{q}{2\pi L}\delta(\omega-v_F\abs{q})
\eea
\subsubsection{Gaussian Fermionic Environment}\label{sec:spinfermion}
Here we consider a system coupled to a fermionic environment, 
\bea
\hat{H}=\hat{H}_s+\sum_k\omega_k\hat{c}_k^\dagger \hat{c}_k+\hat{S}\sum_{k}g_{k}(\hat{c}_{k}^\dagger +\hat{c}_k).
\eea
where $\hat{c}_k$ ($\hat{c}_k^\dagger$) are the fermionic annihilation (creation) operators, $\{\hat{c}_j,\hat{c}_k^\dagger\}=\delta_{jk}$, and we require that $\hat{S}$ is diagonalizable. For the spin-fermion model, $\hat{H}_s=\epsilon_0 \sigma_z+\Delta \sigma_x$ and $\hat{S}=\sigma_z$. 
Starting from Eq. (\ref{eq:continuousPI}), the bath correlation function is different. That is,
\bea
\langle B(\tau)B(\tau') \rangle_B&=&\Big\langle \sum_jg_j(\hat{c}_j^\dagger e^{i\omega_j(\tau-\tau')}+\hat{c}_j e^{-i\omega_j(\tau-\tau')})\sum_k{g_k}(\hat{c}_k^\dagger+\hat{c}_k)\Big\rangle\nonumber\\
&=&\sum_{jk} \langle g_j(\hat{c}_j^\dagger e^{i\omega_j(\tau-\tau')}+\hat{c}_j e^{-i\omega_j(\tau-\tau')}){g_k}(\hat{c}_k^\dagger+\hat{c}_k)\rangle
\nonumber\\
&=&\sum_{j}g_j^2[\langle \hat{c}_j^\dagger e^{i\omega_jt} \hat{c}_j \rangle +\langle \hat{c}_j e^{-i\omega_jt} \hat{c}_j^\dagger\rangle]\nonumber\\
&=&\sum_{j}g_j^2[e^{i\omega_jt}\langle  \hat{n}_F(\omega_j) \rangle +e^{-i\omega_jt} \langle 1-\hat{n}_F(\omega_j)\rangle],
\eea
and taking the continuous limit in addition to using the spectral density $J(\omega)=\pi\sum_j\lambda_j^2\delta(\omega-\omega_j)$, we get 
\bea
\frac{1}{\pi}\int_0^\infty d\omega J(\omega)[e^{i\omega_jt}(  {n}_F(\omega_j) ) +e^{-i\omega_jt} (1-{n}_F(\omega_j))]=\frac{1}{\pi}\int_0^\infty d\omega J(\omega)[\cos(\omega t)+i\sin(\omega t)(2 n_F(\omega)-1)].
\eea
Recall ${n_F}(\omega)=\frac{1}{e^{\beta (\omega-\mu)}+1}$ so $2{n}_F(\omega)-1=\frac{2}{e^{\beta (\omega-\mu)}+1}-1=\frac{e^{\beta (\omega-\mu)/2}-e^{-\beta (\omega-\mu)/2}}{e^{\beta (\omega-\mu)/2}+e^{-\beta (\omega-\mu)/2}}$. So, if we allow $J(\omega)=-J(-\omega)$, this integral becomes half of an even integrand. As a result
\bea
\langle B(t)B(0) \rangle_B=\frac{1}{2\pi}\int_{-\infty}^\infty d\omega J(\omega)\frac{e^{\beta (\omega-\mu)/2}e^{-i\omega t}}{\cosh{\beta (\omega-\mu)/2}}.\label{eq:fermBB}
\eea
Conveniently, this expression Eq. (\ref{eq:fermBB})is only slightly modified compared to the analysis for the spin-boson model. Furthermore, the following relations. 
See Ref. \citenum{jin2007dynamics} with which this result coincides. 
\subsubsection{Gaussian Spin Environment}
Now we consider a system coupled with a diagonalizable coupling operator to a spin bath (again, the generalization to multiple such baths is straightforward if all the coupling operators commute). The Hamiltonian for this model reads \cite{makriss,segal2014two,wang2012dynamics}
\bea
\hat{H}=\hat{H}_s+\frac{1}{2}\sum_i^{N_s}\omega_i\hat{\sigma}_z^i+\hat{S}\frac{1}{2}\sum_i^{N_s}\frac{c_i}{\sqrt{2\omega_i}}\hat{\sigma}_x^i.
\eea
Note that environmental spins are non-interacting. Further, they are coupled to the central system in a manner by which $c_i\propto\frac{1}{\sqrt{N_s}}$ such that these contributions vanish in the thermodynamic limit--this restriction is physically motivated and common in the literature\cite{makriss,segal2014two,wang2012dynamics}. 
To follow we consider concretely the spin-spin bath model $\hat{H}_s=\epsilon \hat\sigma_z+\Delta\hat\sigma_x$ and $\hat{S}=\hat\sigma_z$. 
Even for this simplified model, although a Gaussian nature cannot in general be assumed\cite{link2023non} when considering the thermodynamic limit of the spins ($N_s\to\infty$), one obtains an effective spin-boson picture where the spectral density is renormalized\cite{caldeira} 
\bea
J_{eff}(\omega,\beta)=\tanh{\beta\omega/2}J(\omega).
\eea
As a result, our results from previous models also apply to this case, mapping this problem to the spin-boson model.

\subsection{System propagator in terms of influence functions}\label{subsec:UinI}
Returning from the end of \ref{sec:pi1}, going into Eq. (\ref{eq:UN}) for Gaussian environments, if one writes the propagators in a more compact form (for $k>0$)
\bea
 U_{k,x_{2k}^\pm x_0^\pm}&=& \sum_{x_{2k-1}^\pm} G_{x_{2k}^\pm x_{2k-1}^\pm}\sum_{x_{1}^\pm}\tilde{U}_{k-1,x_{2k-1}^\pm x_{1}^\pm}G_{x_{1}^\pm x_{0}^\pm},\nonumber\\
 F_{x_{2k+1}^\pm x_{2k-1}^\pm}&=&\sum_{x_{2k}^\pm}G_{x_{2k+1}^\pm x_{2k}^\pm}G_{x_{2k}^\pm x_{2k-1}^\pm},
\eea
the auxiliary propagators can be written as
\bea
\tilde{U}_{0,x_{1}^\pm x_{1}^\pm}&=&I_{0,x_{1}^{\pm}}\label{Ut1}\\
\tilde{U}_{1,x_{3}^\pm x_{1}^\pm}&=&F_{x_{3}^\pm,x_{1}^\pm}I_{1,x_{3}^{\pm}x_{1}^{\pm}}I_{0,x_{3}^{\pm}}I_{0,x_{1}^{\pm}}\nonumber\\
&=&\underbrace{F_{x_{3}^\pm x_{1}^\pm}I_{1,x_{3}^{\pm}x_{1}^{\pm}}I_{0,x_{3}^{\pm}}}_{M_{1,x_{3}^\pm x_{1}^\pm}}\tilde{U}_{0,x_{1}^\pm x_{1}^\pm}\\
\tilde{U}_{2,x_{5}^\pm x_{1}^\pm}&=&\sum_{x_{3}^\pm}
F_{x_{5}^\pm x_{3}^\pm}F_{x_{3}^\pm x_{1}^\pm}I_{2, x_{5}^{\pm}x_{1}^{\pm}} I_{1,x_{5}^{\pm}x_{3}^{\pm}}I_{1,x_{3}^{\pm}x_{1}^{\pm}}I_{0,x_{5}^{\pm}}I_{0,x_{3}^{\pm}}I_{0,x_{1}^{\pm}}\nonumber\\
&=&\sum_{x_{3}^\pm}
\underbrace{F_{x_{5}^\pm x_{3}^\pm}I_{1,x_{5}^{\pm}x_{3}^{\pm}}I_{0,x_{5}^{\pm}}}_{M_{1, x_{5}^\pm x_{3}^\pm}}\underbrace{F_{x_{3}^\pm x_{1}^\pm}I_{1,x_{3}^{\pm}x_{1}^{\pm}}I_{0,x_{3}^{\pm}}I_{0,x_{1}^{\pm}}}_{\tilde{U}_{1,x_{3}^\pm x_{1}^\pm}}\nonumber\\
&&+(I_{2,x_{5}^{\pm}x_{1}^{\pm}}-1)\sum_{x_{3}^\pm}
F_{x_{5}^\pm x_{3}^\pm}F_{x_{3}^\pm x_{1}^\pm}I_{1,x_{5}^{\pm}x_{3}^{\pm}}I_{1,x_{3}^{\pm}x_{1}^{\pm}}I_{0,x_{5}^{\pm}}I_{0,x_{3}^{\pm}}I_{0,x_{1}^{\pm}},\label{Ut3}
\eea
where the underbraced $M$ terms and the last line is analogous to the memory matrix in small matrix path integral (SMatPI) methods \cite{smatpi1,smatpi2,Pathsum} (as well as the memory equation approach by Pechukas and co-workers\cite{golosov1999efficient}) and the transfer tensors in the transfer tensor method (TTM)\cite{CaoTTM}.
While we do not use the memory matrix directly in our analysis, we note the form to connect with other known approaches.

We briefly comment on how our decomposition may be related to other known numerical approaches, such as quasiadiabatic propagator path integral (QUAPI), SMatPI, and TTM.
While we emphasize that our analysis is primarily not a numerical tool to perform open quantum system dynamics, one may still wonder how it is connected to other known numerical techniques.
Here, we summarize the key differences:
\q{diffbetmet}{
\begin{enumerate}
\item {\it QUAPI}: 
As explained in our main text, QUAPI assumes $I_k = 1$ for $k > k_\text{max}$. 
This assumption alone does not immediately reveal $\mathbf U$ in terms of $I_k$. If one wrote $\mathbf U$ in terms of $I_k$ using QUAPI without truncation, one would recover the same form of SMatPI without truncation (vide infra).
\item {\it SMatPI}: The SmatPI decomposition writes $\mathbf U$ in terms of influence functions and the bare system propagator. However, in doing so, it loses the time-translational symmetry of the memory functions, which makes the subsequent analysis we perform in this manuscript (see \cref{Appendix B} and \cref{Appendix C F}) inaccessible. Furthermore, the SMatPI approach was never formulated for problems in {\it Classes} 2--4.
\item {\it TTM}: The TTM decomposition, a black-box, data-driven, extrapolative technique, requires projection-free ``numerical'' inputs. In our analysis, we decompose the reduced system dynamical map in terms of the reduced system bare propagator and influence functions. This major insight is not present in TTM.
\end{enumerate}
These differences make our analysis and insight presented in this manuscript unique contributions in this field.}

\subsection{Memory kernel in terms of influence functions} \label{Appendix Bclass1}
As alluded to in \cref{Appendix B}, one can make appropriate substitutions to write $\mathbf U$ in terms of $\mathbf I$ and thereby write $\mathbf K$ in terms of $\mathbf I$.
The direct and explicit relationship between $\mathbf K$ and $\mathbf I$, to our knowledge, has not been explicitly written for any {\it Classes} beyond numerical approaches using INFPI to obtain projection-free inputs for the NZ equation.
We here focus on how to show this for {\it Class 1}. 

For {\it Class 1}, we unravel Eq.~(\ref{eq:KKK1}) and obtain
\bea
(\bm{\mathcal{K}}_{0})_{ik}=\frac{1}{\Delta t^2}\Big[\sum_jG_{ij}I^{0}_{j}G_{jk}+(\frac{i}{\hbar}\mathcal{L}_s\Delta t-1)_{ik}\Big].
\eea
Next, from \cref{eq:appendixK1}, 
\bea
(\bm{\mathcal{K}}_{1})_{im}=\frac{1}{\Delta t^2}\Big[\sum_jG_{ij}\sum_kF_{jk}I_{1,jk}I_{0,j}I_{0,k}G_{km}-\sum_jG_{ij}I_{0,j}\sum_kG_{jk}\sum_lG_{kl}I_{0,l}G_{lm}\Big].
\eea
Similarly by making appropriate substitutions into $\mathbf U_1$, $\mathbf U_2$, and $\mathbf U_3$ in \cref{eq:appendixK2},
\bea
(\bm{\mathcal{K}}_2)_{ip}&=&\frac{1}{\Delta t^2}\Big[-\sum_jG_{ij}I_{0,j}\sum_kF_{jk}I_{1,jk}I_{0,k} \sum_nF_{kn} I_{0,n} G_{np}+\sum_jG_{ij}I_{0,j}\sum_k F_{jk}I_{0,k}\sum_n F_{kn}I_{0,n}G_{np}\nonumber\\
&&-\sum_jG_{ij}I_{0,j}\sum_kF_{jk} \sum_nF_{kn} I_{1,kn}I_{0,k}I_{0,n}  G_{np}+
\sum_{jkn}
G_{ij}F_{jk}F_{kn}I_{2,jn} I_{1,jk}I_{1,kn}I_{0,j}I_{0,k}I_{0,n}G_{np}.\label{k2}
\eea
We again factorize as follows
\bea
I_{2,jn} I_{1,jk}I_{1,kn}=(I_{2,jn}-1)I_{1,jk}I_{1,kn}+\underbrace{I_{1,jk}+I_{1,kn}-1}+(I_{1,jk}-1)(I_{1,kn}-1).
\eea
The underbraced terms cancel with the first three terms in Eq. (\ref{k2}) and we are left with
\bea
(\bm{\mathcal{K}}_2)_{ip}&=&\sum_{jkn}
G_{ij}F_{jk}F_{kn}[(I_{2,jn}-1)I_{1,jk}I_{1,kn}+(I_{1,jk}-1)(I_{1,kn}-1)]I_{0,j}I_{0,k}I_{0,n}G_{np}.
\eea

One can write down higher-order terms similarly.
By making substitutions to \cref{eq:appendixK3}, 
\bea
(\bm{\mathcal{K}}_{3})_{il}&=&\sum_{jknp}
G_{ij}F_{jk}F_{kn}F_{np}[I_{3,jp}I_{2,jn}I_{2,kp}I_{1,jk}I_{1,kn}I_{1,np}+I_{1,jk}-1+I_{1,kn}-I_{2,jn}I_{1,jk}I_{1,kn}\nonumber\\&&-I_{1,jk}I_{1,np}+I_{1,np}-I_{2,kp}I_{1,kn}I_{1,np}]I_{0,j}I_{0,k}I_{0,n}I_{0,p}G_{pl}. \label{K40}
\eea
We factorize as follows
\bea
I_{3,jp}I_{2,jn}I_{2,kp}I_{1,jk}I_{1,kn}I_{1,np}&=&(I_{3,jp}-1)I_{2,jn}I_{2,kp}I_{1,jk}I_{1,kn}I_{1,np}+I_{2,jn}I_{2,kp}I_{1,jk}I_{1,kn}I_{1,np}.
\eea
For now notice that
\bea
&&I_{2,jn}I_{2,kp}I_{1,jk}I_{1,kn}I_{1,np}-I_{2,jn}I_{1,jk}I_{1,kn}-I_{2,kp}I_{1,kn}I_{1,np}=I_{1,kn}(I_{2,jn}I_{2,kp}I_{1,jk}I_{1,np}-I_{2,jn}I_{1,jk}-I_{2,kp}I_{1,np})\nonumber\\
&=&I_{1,kn}[(I_{2,jn}-1)(I_{2,kp}-1)I_{1,jk}I_{1,np}+I_{1,jk}I_{1,np}-(I_{2,jn}-1)I_{1,jk}-I_{1,jk}-(I_{2,kp}-1)I_{1,np}-I_{1,np})\nonumber\\
&&(I_{2,kp}-1)I_{1,jk}I_{1,np}+(I_{2,jn}-1)I_{1,jk}I_{1,np}]
\eea
because 
\bea
I_{2,jn}I_{2,kp}I_{1,jk}I_{1,np}&=&I_{2,kp}I_{1,jk}I_{1,np}+(I_{2,jn}-1)I_{2,kp}I_{1,jk}I_{1,np}\nonumber\\
&=&(I_{2,kp}-1)I_{1,jk}I_{1,np}+(I_{2,jn}-1)I_{1,jk}I_{1,np}+I_{1,jk}I_{1,np}\nonumber\\&&+(I_{2,jn}-1)I_{2,kp}I_{1,jk}I_{1,np}+(I_{2,jn}-1)(I_{2,kp}-1)I_{1,jk}I_{1,np}
\eea
note there
\bea
&&(I_{2,jn}-1)(I_{2,kp}-1)I_{1,jk}I_{1,np}+(I_{2,kp}-1)I_{1,jk}I_{1,np}+(I_{2,jn}-1)I_{1,jk}I_{1,np}-(I_{2,jn}-1)I_{1,jk}-(I_{2,kp}-1)I_{1,np}\nonumber\\
=&&(I_{2,jn}-1)(I_{2,kp}-1)I_{1,jk}I_{1,np}+(I_{2,kp}-1)(I_{1,jk}-1)I_{1,np}+(I_{2,jn}-1)(I_{1,np}-1)I_{1,jk}
\eea
and also that
\bea
I_{1,kn}[I_{1,jk}I_{1,np}-I_{1,jk}-I_{1,np}]+I_{1,jk}-1+I_{1,kn}-I_{1,jk}I_{1,np}+I_{1,np}=(I_{1,jk}-1)(I_{1,kn}-1)(I_{1,np}-1).
\eea
These resulting identities then group terms in Eq. (\ref{K40}) to give Eq. (\ref{eq:K3}) in the main text.

Now for $\bm{\mathcal{K}}_{4}$ (we use $\tilde{I}_{i,jk}=(I_{i,jk}-1)$):
\bea
\bm{\mathcal{K}}_{4,im}&=&\frac{1}{\Delta t^2}\sum_{jknpl}
G_{ij}F_{jk}F_{kn}F_{np}F_{pl}I_{0,j}I_{0,k}I_{0,n}I_{0,p}I_{0,l}G_{lm}\Big\{\nonumber\\
&&\tilde{I}_{4,jl}I_{1,jk}I_{2,jn}I_{3,jp}I_{1,kn}I_{2,kp}I_{3,kl}I_{1,np}I_{2,nl}I_{1,pl}\nonumber\\&&+
\tilde{I}_{3,jp}I_{1,jk}I_{2,jn}I_{1,kn}I_{2,kp}I_{1,np}\tilde{I}_{3,kl}I_{2,nl}I_{1,pl}+
\tilde{I}_{3,jp}I_{1,jk}I_{2,jn}I_{1,kn}I_{2,kp}I_{1,np}\tilde{I}_{2,nl}I_{1,pl}\nonumber\\
&&+
\tilde{I}_{3,jp}I_{1,jk}I_{2,jn}I_{1,kn}I_{2,kp}I_{1,np}\tilde{I}_{1,pl}+
\tilde{I}_{2,jn}I_{1,jk}I_{1,kn}\tilde{I}_{3,kl}I_{2,kp}I_{1,np}I_{2,nl}I_{1,pl}+
\tilde{I}_{2,jn}I_{1,jk}I_{1,kn}\tilde{I}_{2,kp}I_{1,np}\tilde{I}_{2,nl}I_{1,pl}\nonumber\\
&&+
\tilde{I}_{2,jn}I_{1,jk}I_{1,kn}\tilde{I}_{2,kp}I_{1,np}\tilde{I}_{1,pl}+
\tilde{I}_{2,jn}I_{1,jk}I_{1,kn}\tilde{I}_{2,nl}I_{1,np}I_{1,pl}+
\tilde{I}_{2,jn}I_{1,jk}I_{1,kn}\tilde{I}_{1,np}\tilde{I}_{1,pl}\nonumber\\
&&+
\tilde{I}_{1,jk}\tilde{I}_{3,kl}I_{1,kn}I_{2,kp}I_{1,np}I_{2,nl}I_{1,pl}+
\tilde{I}_{1,jk}\tilde{I}_{2,kp}I_{1,kn}I_{1,np}\tilde{I}_{2,nl}I_{1,pl}+
\tilde{I}_{1,jk}\tilde{I}_{2,kp}I_{1,kn}I_{1,np}\tilde{I}_{1,pl}\nonumber\\
&&+
\tilde{I}_{1,jk}\tilde{I}_{1,kn}\tilde{I}_{2,nl}I_{1,np}I_{1,pl}+
\tilde{I}_{1,jk}\tilde{I}_{1,kn}\tilde{I}_{1,np}\tilde{I}_{1,pl}
\Big\}.
\eea
The number of terms grows combinatorially with the number of timesteps $N$. In the next section, we devise a general scheme to build memory kernels at arbitrary time $N\Delta t$ based on Dyck diagrams.
\subsection{Nakajima-Zwanzig Dyck-Diagrammatic Formalism}\label{sec:Dyck}
The structure of terms in $\bm{\mathcal{K}}_N$ via the decomposition through $\{\mathbf{{I}}_k\}$ has the following properties. The number of terms in $\bm{\mathcal{K}}_N$ is given by Catalan's numbers\cite{catalanbook,catalanoies} $C_N=\frac{1}{N+1}{{2N}\choose{N}}$, as was pointed out in Ref. \citenum{tree-smatpi}. More importantly, the terms in $\bm{\mathcal{K}}_N$ are given by Dyck path diagrams \cite{catalanbook,catalanoies} (of which there are $C_N$ paths). A Dyck path is a sequence of up-steps and down-steps that start and end on the $x$-axis but never go below. The order of a Dyck path is given by the number of up-steps, which must equal the number of down-steps. The study of Dyck paths is at the root of combinatorial mathematics\cite{vilenkin1971combinatorics}, and algorithms already exist to generate all Dyck paths of any order.\cite{dyckalg1,dyckalg2} 
Alternatively, one can also generate all $P^R(N,k)=k^N$ (sometimes also called k-tuples) possible permutations (repetitions allowed, since here $k=2$ accounting for up or down steps we then have  $P^R(2,2)=4,P^R(4,2)=16,P^R(6,2)=64$, and so on) of $\nearrow$ and $\searrow$ and then filter (i.e., such path must have equal amounts of $\nearrow$ and $\searrow$ and at any point the path must not go below the $x$-axis and so on) for an order $2N$ Dyck path. To obtain the terms in $\bm{\mathcal{K}}_N$ from a particular Dyck path, we map a Dyck path into an influence functional diagram based on the following rules:
\begin{enumerate}
    \item The Dyck path is made dashed.
    \item Now, one draws all possible triangles in each segment of a Dyck path, but none could be as tall as any dashed peak. This step is drawn in solid. A segment of a Dyck path is defined as the sub-path of a Dyck path which starts and ends at $x=0$ (the full path itself is not counted). A peak is when an up-step meets a down-step in that order.
    \item If any dashed down path sequence does not reach $x=0$, one continues to draw it until it does, with dashed lines.
    \item One rounds the vertices.
    \item Dashed correlations connecting points $i$ and $j$ are $I_{ij}-1$, while solid correlations are $I_{ij}$. Further terms can be read immediately from the diagrams.
\end{enumerate}
An example of this algorithm is shown in Fig. (\ref{Algorithm}).
\begin{figure}[hbt!]
    \centering
    \includegraphics[width=0.9\columnwidth]{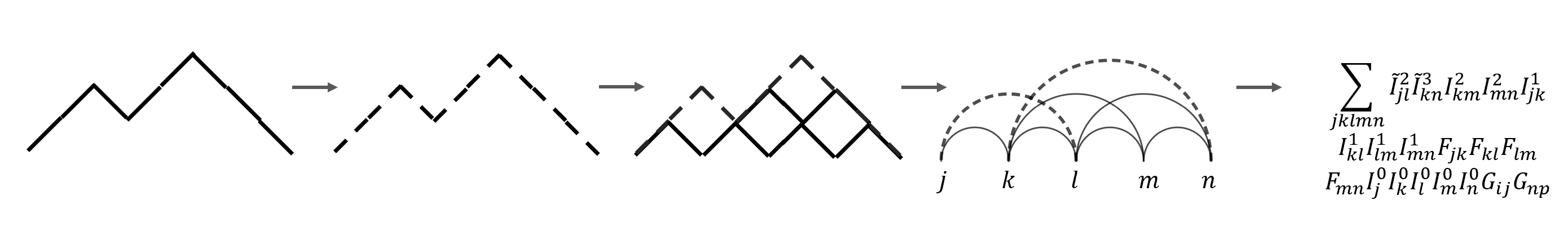}
    \caption{Visual schematic of the algorithm to convert Dyck diagrams into terms in $\bm{\mathcal{K}}$.}
    \label{Algorithm}
\end{figure}

We furthermore enumerate some properties of Dyck paths, which therefore unveils the general properties of $\bm{\mathcal{K}}_N$:
\begin{enumerate}
\item The set of all Dyck paths of order $n$ has Narayana's number, $Nr(n,k)$, the number of paths with $k$ peaks. The number of peaks is the multiplicity of $\tilde{I}$ in the term. $Nr(n,k)$ is given by
    \bea
    Nr(n,k)=\frac{1}{n}{{n}\choose{k}}{{n}\choose{k-1}}
    \eea
    It is not difficult to show that $\sum_k Nr(n,k)=C_n$.
\item The number of Dyck paths in the set of all Dyck paths of order $n$ having $k$ number of segments is given by Catalan's triangle $C(n,k)$:
    \bea
        C(n,k)=\frac{n-k+1}{n+1}{{n+k}\choose{k}}.
    \eea
    In particular, if $n=k$, $C(n,n)$ is the $n$th Catalan's number $C_n$.
\item The number of Dyck paths in the set of all Dyck paths of order $n$ having no hills (a segment of height $1$), that is, those with no $\tilde{I}_{1}$ term, is given by the Fine numbers $F_n$ with generating function $F(x)$\cite{DEUTSCH2001241}:
    \bea
        F(x)=\sum_{n\geq0}F_nx^n=\frac{1-\sqrt{1-4x}}{x(3-\sqrt{1-4x})}.
    \eea
    This distribution also describes (1) Dyck paths where the leftmost peak is of even height,\cite{DEUTSCH2001241} and (2) the number of even returns to the horizontal axis (valley of height 0)\cite{DEUTSCH2001241}.
    \item The number of Dyck paths in the set of all Dyck paths of order $2n+2m$ with $m$ returns to the horizontal axis is given by the Ballot numbers $B(n,m)$
    \bea
    B(n,m)=\frac{m-n}{m+n}{{m+n}\choose{n}}, \ \ \ (m,n)\neq(0,0)
    \eea
\item The number of Dyck paths in the set of all Dyck paths of order $n$ with $k$ 
occurrences of different $\nearrow\searrow\nearrow$ is given by $T(n,k)$, where $T(n,k)$ is related to Motzkin's number $M(k)$ via\cite{SUN2004177} 
    \bea
    T(n+1,k)={{n}\choose{k}}M(n-k).
    \eea
    Particularly, the generator of Motzkin's number $M(k)$, $m(x)=\sum^\infty_0 M(n)x^n$, satisfies
    \bea
    x^2m(x)^2+(x-1)m(x)+1=0.
    \eea
\item The number of Dyck paths in the set of all Dyck paths of order $n$ whose peaks' height total to $k$ can be read off from Triangle A094449.\cite{oeis}
\end{enumerate}
Further properties of the Dyck paths have been enumerated in great detail in the On-Line Encyclopedia of Integer Sequences\cite{oeis}.
To construct the memory kernels via this approach, we can represent Dyck paths as Dyck Words. $\nearrow$ becomes $1$ and $\searrow$ becomes $0$. Hence, the Dyck path in Fig. (\ref{Algorithm}) is $11011000$.  
By representing each path by a bit string, we can write an efficient code that enumerates all possible Dyck paths of order $N$.

\subsection{Details on Computing $J(\omega)$ from $\rho(t)$}\label{Appendix C F}
For Gaussian environments (here, we concretely consider spin-boson, but extensions to other environments are straightforward; as we discuss further in the next appendices, we provide more details on the inverse extraction procedure that goes from reduced system dynamics, $\rho(t)$, to the bath spectral density, $J(\omega)$. We show that such a map is nearly bijective in that any aspects of $J(\omega)$ that affect the reduced system dynamics can be obtained by analyzing $\rho(t)$. This construction directly connects with Hamiltonian learning literature, as pointed out in our main text.

To proceed, we recall
\bea
\rho\leftrightarrow\mathbf{U}\leftrightarrow\bm{\mathcal{K}}\leftrightarrow\mathbf{I}\leftrightarrow\eta\leftrightarrow J(\omega).
\eea
We will explain step-by-step how to go from left to right.
First, to obtain $\mathbf{U}_N$ from $\rho_N$ note that $\rho_N=\mathbf{U}_N\rho_0$, where $\rho_N$ is a vector in Liouville space with a dimension of $N_L$. $\mathbf{U}_N$ is then a $N_{L} \times N_{L}$ matrix. With $N_L$ linearly-independent trajectories (obtained experimentally or computationally), we can uniquely determine $\mathbf U_N$. We then stack them to make the matrix
\bea
\boldsymbol{\mathrm{P}}_N=[\rho_N^{(1)}|\rho_N^{(2)}|\dots|\rho_N^{(N_{L})}].
\eea
With this definition, we arrive at a simple linear equation to solve,
\begin{equation}
\boldsymbol{\mathrm{P}}_N
=
\mathbf{U}_N
\boldsymbol{\mathrm{P}}_0
\end{equation}
By inverting $\boldsymbol{\mathrm{P}}_0$, we obtain
\bea
\mathbf{U}_N=\boldsymbol{\mathrm{P}}_N\boldsymbol{\mathrm{P}}_0^{-1}.
\eea
$\boldsymbol{\mathrm{P}}_0$ is invertible because each trajectory is generated from a linearly independent initial condition.
While we only considered noise-free data inputs, if the data is noisy, one can use more trajectories than $N_L$ (i.e., $\boldsymbol{\mathrm{P}}_N$ and $\boldsymbol{\mathrm{P}}_0$ become a fat matrix) and perform the Moore-Penrose pseudoinverse, which could help mitigate noise.

To obtain $\bm{\mathcal{K}}$ from $\mathbf{U}$ we decompose the latter iteratively, in the manner presented in the main text Eq. (\ref{mEMTTIM}) and e.g., Ref. \citenum{CaoTTM}.

Next, we move to obtain $\mathbf{I}$ from $\bm{\mathcal{K}}$. First, we define $\tilde{\bm{\mathcal{K}}}_N$ which is defined through $\mathbf{G}\tilde{\bm{\mathcal{K}}}_N\mathbf{G}={\bm{\mathcal{K}}}_N$. 
From $\tilde{\bm{\mathcal{K}}}_0$, we have 
\begin{equation}
\Delta t^2(\tilde{\bm{\mathcal{K}}}_0+\mathbf G^{-1}\mathbf L\mathbf G^{-1})_{ii}=I_{0,i}
\end{equation}
Similarly, we have
\bea
\tilde{{\mathcal{K}}}_{1,jk}=
\frac{1}{\Delta t^2}I_{0,j}F_{jk}\tilde{I}_{1,jk}I_{0,k}&\Rightarrow& \tilde{I}_{1,jk}=\Delta t^2\frac{\tilde{{\mathcal{K}}}_{1,jk}}{F_{jk}\tilde{U}_{0,jj}\tilde{U}_{0,kk}}.
\eea
For $N=2$, we have
\bea
\tilde{{\mathcal{K}}}_{2,jn}&=&\tilde{I}_{2,jn}\frac{1}{\Delta t^2}\sum_{k}
F_{jk}F_{kn}I_{1,jk}I_{1,kn}
I_{0,j}I_{0,k}I_{0,n}+\frac{1}{\Delta t^2}\sum_{k}
F_{jk}F_{kn}
\tilde{I}_{1,jk}\tilde{I}_{1,kn}I_{0,j}I_{0,k}I_{0,n},
\eea
which leads to
\bea
\tilde{I}_{2,jn}&=&\frac{\Delta t^2\tilde{{\mathcal{K}}}_{2,jn}-\sum_{k}
F_{jk}F_{kn}
\tilde{I}_{1,jk}\tilde{I}_{1,kn}I_{0,j}I_{0,k}I_{0,n}}{\sum_{k}
F_{jk}F_{kn}I_{1,jk}I_{1,kn}
I_{0,j}I_{0,k}I_{0,n}}.\label{I2jn}
\eea
This procedure is, in fact, diagrammatically generalizable--its diagrammatic equation is shown in Fig. (\ref{Idiagram}). One can generate all Dyck diagrams (which can be translated into a particular term in $\bm{\mathcal{K}}_N$) of order $N$. From the sum of all diagrams, one moves all diagrams except the crest term to the side of $\bm{\mathcal{K}}_N$. The crest term corresponds to the diagram that contains $\tilde{\mathbf{{I}}}_N$. Then, one divides both sides of the equation with the rest of the terms, which then yields $\tilde{\mathbf{{I}}}_N$. We note the computational costs of the procedure scale (super)combinatorially with the Dyck order $N$ as one must compute Catalan's numbers $C_N=\frac{1}{N+1}{{2N}\choose{N}}$ of diagrams in $\bm{\mathcal{K}}_N$, each with about $N-1$ of contractions. However, there are ample opportunities to improve this procedure numerically, for example, by including specific classes of the Dyck diagram (see \ref{sec:Dyck}).
\begin{figure}[hbt!]
    \centering
    \includegraphics[width=0.5\columnwidth]{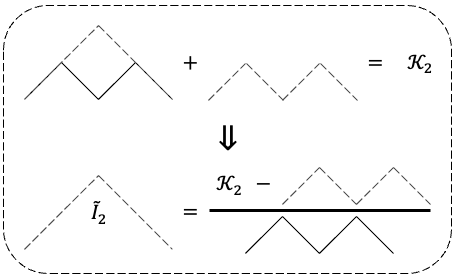}
    \caption{Visual schematic of the inversion diagram for $\tilde{\mathbf{{I}}}_2$}
    \label{Idiagram}
\end{figure}

Now to obtain $\eta$ from $\mathbf{I}$, we start by taking the logarithm of \cref{eq:Ikk'},
\bea
\ln{I_{k'-k,x_{k}^{\pm}x_{k'}^{\pm}}}=-(x_k^+-x_k^-)(\eta_{kk'}x_{k'}^+-\eta_{kk'}^*x_{k'}^-).
\eea
Here, we consider the spin-boson model with $\sigma_z$ coupling, so ${x}^\pm\in\text{pair}\{+1,-1\}$. In this example, we have
\bea
\ln{I_{(k'-k)[1,-1][1,-1]}}&=&4\Re\eta_{kk'}\\
\ln{I_{(k'-k)[1,-1][1,1]}}&=&4i\Im\eta_{kk'}.
\eea
We can also readily obtain $\eta_{kk}$ from $I_{0}$ through a similar manner. 

Lastly, to obtain the spectral density, $J(\omega)$, we recall that (for clarity $k-k'\to\Delta k$)
\bea
\eta_{\Delta k}&=&\int_{-\infty}^{\infty}d\omega \mathcal F(\omega)e^{-i\omega\Delta t \Delta k}.\label{etaJ}
\eea
where
\bea
\mathcal F(\omega)=\frac{2}{\pi}\frac{J(\omega)}{\omega^2}\frac{\exp{\beta\hbar\omega/2}}{\sinh{\beta\hbar\omega/2}}\sin^2{(\omega\Delta t/2)}\label{A83}
\eea
Note, in practice, we construct $\eta_{\Delta k}$ from $I_{kk'}$ for $\Delta k >0$ and take the hermitian conjugate of $\eta_{\Delta k}$ for $\Delta k<0$. We use $I_{kk}$ for $\eta_0$.
To extract $\mathcal F(\omega)$ with the knowledge of $\eta_{\Delta k}$ we perform a Fourier Transform
\bea
\mathcal F(\omega)=\frac{1}{2\pi}\int_{-\infty}^\infty d{t}\eta_{\Delta k}e^{i\omega \Delta t\Delta k},
\eea
which, in practice, is performed via a discrete Fourier transform
\begin{equation}
\mathcal F(\omega)=\frac{\Delta t}{2\pi}\sum_{\Delta k=-\Delta k_{\text{max}}}^{\Delta k_{\text{max}}} \eta_{\Delta k}e^{i\omega \Delta t\Delta k}.
\label{A85}
\end{equation}
This immediately gives the spectral density through
\begin{equation}
J(\omega)=\mathcal F(\omega)\frac{\pi\omega^2\sinh{\beta\hbar\omega/2}}{2\sin^2{(\omega\Delta t/2)}\exp{\beta\hbar\omega/2}}.
\label{A86}
\end{equation}
\q{invcorr}{
One must pay special attention to frequencies with $\omega\Delta t /2=n\pi$ for integer $n$ as the denominator in Eq. (\ref{A86}) is zero and the expression diverges. However, this is not a problem because, physically, the spectral density at such nodal $\omega$ values does not alter the reduced system dynamics. This is because they do not contribute to $\mathcal F(\omega)$ as shown in Eq. (\ref{A83}), and can thus be seen as the null kernel of the map $J(\omega)\to\eta$. To see this, we substitute these frequencies to Eq. (\ref{A85})
\bea
\mathcal F(\frac{2n\pi}{\Delta t})&=&\frac{\Delta t}{2\pi}\sum_{\Delta k=-\Delta k_{\text{max}}}^{\Delta k_{\text{max}}} \eta_{\Delta k}e^{i2n\pi\Delta k}\nonumber\\
&=&\frac{\Delta t}{2\pi}\sum_{\Delta k=-\Delta k_{\text{max}}}^{\Delta k_{\text{max}}} \eta_{\Delta k}\underbrace{\cos{2n\pi\Delta k}+i\sin{2n\pi\Delta k}}_{1}\nonumber\\
&=&\frac{\Delta t}{\pi}\sum_{\Delta k=0}^{\Delta k_{\text{max}}} \Re\eta_{\Delta k}.
\eea
However, as we substitute the nodal frequencies to Eq. (\ref{A83}) we then see that indeed the real part of $F(\frac{2n\pi}{\Delta t})$ also vanish identically:
\bea
\mathcal F(\frac{2n\pi}{\Delta t})=\frac{2}{\pi}\frac{J(\omega)}{\omega^2}\frac{\exp{\beta\hbar\omega/2}}{\sinh{\beta\hbar\omega/2}}\underbrace{\sin^2{{2n\pi}}}_{0}=0.
\eea
}

Another case to consider is the pure dephasing limit where $\hat{H}_S$ is diagonal and commutes with $\hat{H}_I$.
In this case, only the real part of $\eta$ is available,
\bea
\Re \eta_{\Delta k}&=&\Re \int_{-\infty}^{\infty}d\omega \mathcal F(\omega)e^{-i\omega\Delta t \Delta k}\nonumber\\
&=&\int_{-\infty}^{\infty}d\omega \mathcal F(\omega)\cos{\omega \Delta t\Delta k}.
\eea
We can then perform the inverse cosine transform to obtain $\mathcal F(\omega)$,
\bea
\mathcal F(\omega)= \frac{1}{\pi}\int_{-\infty}^{\infty}d\omega \mathcal \Re \eta_{\Delta k}\cos{\omega \Delta t\Delta k}.
\eea
While the results in the main text concern sensing $J(\omega)$ of a bosonic environment, it is also possible to obtain $J(\omega)$ of a fermionic environment.
For example, we employ the inversion procedure within the model in \ref{sec:spinfermion} in Fig. \ref{fig:fermJw}, with the spectral density following that of Ref. \citenum{ng2023real}, and find promising results.
\begin{figure}[hbt!]
    \centering
\includegraphics[width=0.5\columnwidth]{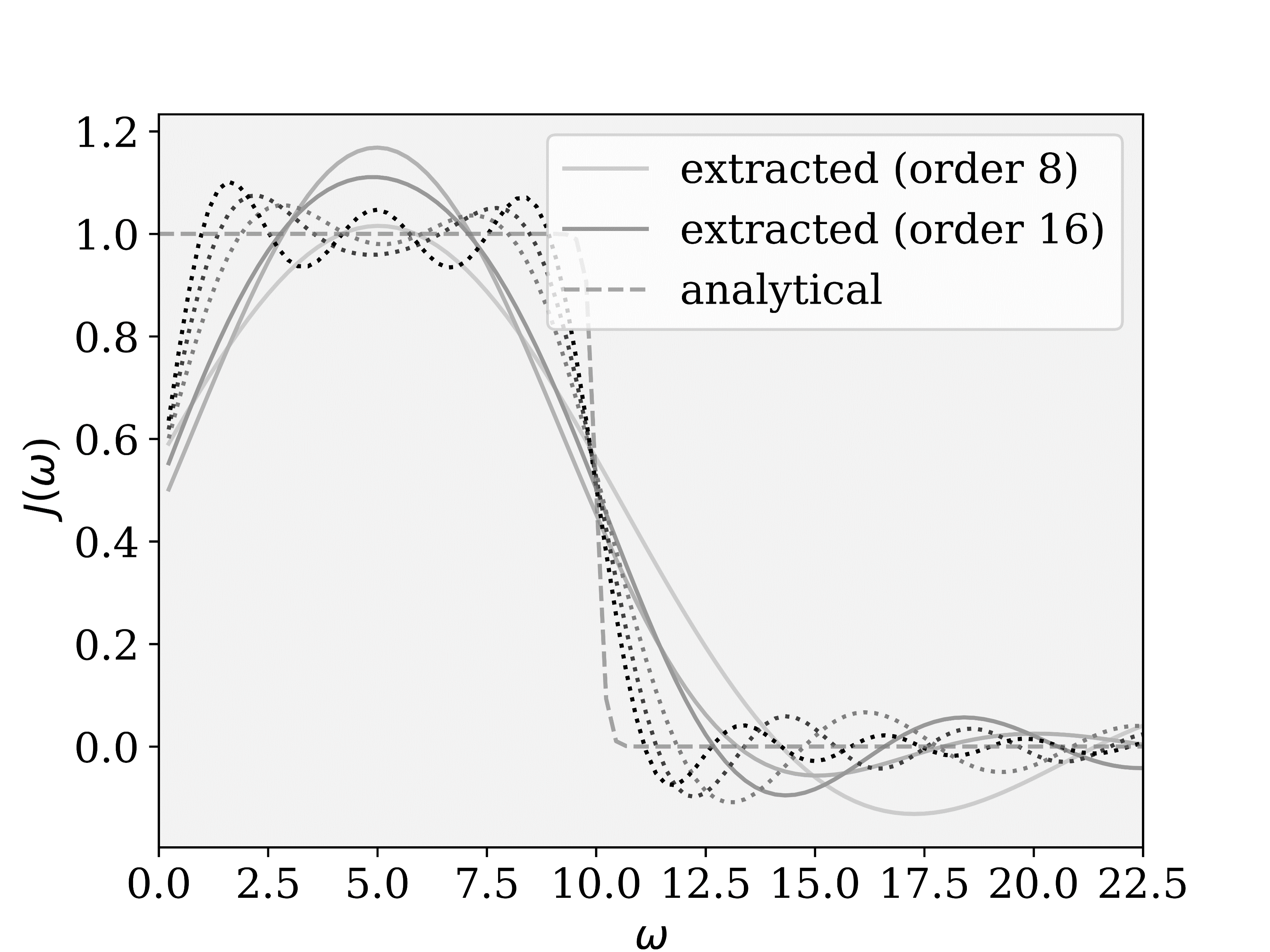}
    \caption{
    Fermionic Bath spectral densities extracted through the Dyck diagrammatic method with increasing truncation orders (from white to black colors, $8,12,16,20,30,$ and $40$ in sequence) compared to exact spectral densities (dashed), see \ref{Appendix C F}. Note that for orders $20, 30,$ and $40$ (dotted), we directly proceed from $\mathbf{I}\to\eta\to J(\omega)$. Here, the fermionic spectral density is defined by defined as a flat band with a smooth edge, adopted from Ref. \citenum{ng2023real}, $J(\omega)=\frac{\gamma}{(1+e^{\nu(\omega-\omega_c)}(1+e^{-\nu(\omega+\omega_c)}}$. 
    Parameters used are: $\Delta=1$ (other parameters are expressed relative to $\Delta$), $\epsilon=0$, $\beta=50$, $\Delta t=0.05$, $\omega_c=10$, $\nu=1/10$, $\mu=0$ and $\gamma=1$.}
    \label{fig:fermJw}
\end{figure}
\subsection{Inversion procedure with multiple environments}\label{sec:invmul}
If we consider now a central system coupled to multiple environments (with simultaneously diagonalizable coupling operators), we will find that 
\bea
I_{k'-k,x_k^\pm x_{k'}^\pm}\to\prod_j I_{k'-k,x_k^\pm x_{k'}^\pm}^j.
\eea
Consequently, we can perform the inversion procedure as in \ref{Appendix C F} but only up to Eq. \ref{I2jn}. This is because what we obtain is 
\bea
I_{k'-k,x_{k}^{\pm} x_{k'}^{\pm}}=\prod_j \exp{-(x_{j,k}^+-x_{j,k}^-)(\eta_{kk'}^jx_{j,k'}^+-\eta_{kk'}^{j,*}x_{j,k'}^-)},
\eea
from which we cannot obtain unique values for each $\eta$. This underdetermined problem is otherwise solved if we obtain the IFs of each bath sequentially. For example, if there are two baths, we can obtain the spectral density of bath 2 from the dynamics of the system influenced by both baths 1 and 2, \textit{if} we have access to the dynamics of the system influenced by bath 1 only.

\subsection{Driven open quantum system dynamics}\label{sec:gaussian}
Here, we consider the explicit time dependence in the system Hamiltonian. We start by inspecting the INFPI formalism without assuming any time-translational invariance of any tensors:
\bea
 U_{k,0,x_{2k}^\pm x_0^\pm}&=& \sum_{x_{2k-1}^\pm} G_{2k-1,x_{2k}^\pm x_{2k-1}^\pm}\sum_{x_{1}^\pm}\tilde{U}_{k-1,0,x_{2k-1}^\pm x_{1}^\pm}G_{0,x_{1}^\pm x_{0}^\pm},\nonumber\\
 F_{2k-1,x_{2k+1}^\pm x_{2k-1}^\pm}&=&\sum_{x_{2k}^\pm}G_{2k,x_{2k+1}^\pm x_{2k}^\pm}G_{2k-1,x_{2k}^\pm x_{2k-1}^\pm},
\eea
with $G_{m,x^\pm_{m+1}x^\pm_{m}}=\langle x^+_{m+1}|e^{-\frac{i\hat{H}_{s}(t_m/2) \Delta t}{2}}| x^+_{m}\rangle\langle x^-_{m}|e^{\frac{i\hat{H}_{s}(t_m/2) \Delta t}{2}}| x^-_{m+1}\rangle$. There, $\hat{H}_{s}(t_m)$ is the system  Hamiltonian at time $t_m = m \Delta t$. Hence,

\bea
\tilde{U}_{0,0,x_{1}^\pm x_{1}^\pm}&=&I_{0,x_{1}^{\pm}}\label{Udt1}\\
\tilde{U}_{1,0,x_{3}^\pm x_{1}^\pm}&=&F_{1,x_{3}^\pm x_{1}^\pm}I_{1,x_{3}^{\pm}x_{1}^{\pm}}I_{0,x_{3}^{\pm}}I_{0,x_{1}^{\pm}}\nonumber\\
\tilde{U}_{2,0,x_{5}^\pm x_{1}^\pm}&=&\sum_{x_{3}^\pm}
F_{3,x_{5}^\pm x_{3}^\pm}F_{1,x_{3}^\pm x_{1}^\pm}I_{2,x_{5}^{\pm}x_{1}^{\pm}} I_{1,x_{5}^{\pm}x_{3}^{\pm}}I_{1,x_{3}^{\pm}x_{1}^{\pm}}I_{0,x_{5}^{\pm}}I_{0,x_{3}^{\pm}}I_{0,x_{1}^{\pm}}\label{Udt3}.
\eea
Similarly, the memory kernels are no longer time-translationally invariant. Consider
\bea
\bm{\mathcal{K}}_{0,0}&=&\mathbf{U}_{1,0}-\mathbf{L}_0\\
\bm{\mathcal{K}}_{1,0}&=&\mathbf{U}_{2,0}-\mathbf{L}_1\mathbf{U}_{1,0}-\bm{\mathcal{K}}_{1,1}\mathbf{U}_{1,0}
\eea
where $\mathbf L_n \equiv (\mathbf{1}-\frac{i}{\hbar}\mathcal{L}_{S,n}\Delta t)$ with
$\mathcal{L}_{S,n} \bullet\equiv[\hat{H}_S(n \Delta t),\bullet]$. We also have
\bea
\bm{\mathcal{K}}_{1,1}&=&\mathbf{U}_{2,1}-\mathbf{L}_1
\eea
Hence,
\bea
\bm{\mathcal{K}}_{1,0}&=&\mathbf{U}_{2,0}-\mathbf{U}_{2,1}\mathbf{U}_{1,0}.
\eea
This allows us to write $\bm{\mathcal{K}}_{1,0}$ in terms of $\mathbf I$:
\bea
(\bm{\mathcal{K}}_{1,0})_{im}&=&\frac{1}{\Delta t^2}\Big[\sum_jG_{3,ij}\sum_kF_{1,jk}I_{1,jk}I_{0,j}I_{0,k}G_{0,km}-\sum_jG_{3,ij}I_{0,j}\sum_kG_{1,jk}\sum_lG_{1,kl}I_{0,l}G_{0,lm}\Big]\nonumber\\
&=&\frac{1}{\Delta t^2}\Big[\sum_jG_{3,ij}\sum_kF_{1,jk}I_{1,jk}I_{0,j}I_{0,k}G_{0,km}-\sum_jG_{3,ij}I_{0,j}\sum_kF_{1,jk}I_{0,k}G_{0,km}\Big]\nonumber\\
&=&\frac{1}{\Delta t^2}\Big[\sum_{jk}G_{3,ij}F_{1,jk}(I_{1,jk}-1)I_{0,j}I_{0,k}G_{0,km}\Big].
\eea
Inspecting equations for later time $\mathcal K$, we observed that the structure of the $\bm{\mathcal{K}}$ remains largely unchanged except for including time indices for the bare system propagators. In particular, we collect all the terms except the time-dependent ones (only the bare system propagators) as a tensor, which is time-translational. That is,
\bea
\bm{\mathcal{K}}_{1+s,s;im}&=&\frac{1}{\Delta t^2}\Big[\sum_{jk}\underbrace{G_{s+3,ij}F_{s+1,jk}G_{s,km}}_{P_{ijkm}^{s+2,s}}T_{1,jk}\Big],\\
\bm{\mathcal{K}}_{2+s,s;ip}&=&\frac{1}{\Delta t^2}\sum_{jkn}
\underbrace{G_{s+5,ij}F_{s+3,jk}F_{s+1,kn}G_{s,np}}_{P_{ijknp}^{s+3,s}}T_{2,jkn},
\eea
where $T_{1,jk}=(I_{1,jk}-1)I_{0,j}I_{0,k}$, $T_{2,jkn}= (\tilde{I}_{2,jn}I_{1,jk}I_{1,kn}
+\tilde{I}_{1,jk}\tilde{I}_{1,kn})I_{0,j}I_{0,k}I_{0,n}$. 
\q{PT2}{As mentioned in the main text, $\mathbf T$ takes a similar form as that of the process tensor,~\cite{jorgensen2019, Jorgensen2020Nov}
but it is distinct in that it directly represents $\mathbf K$, as opposed to $\mathbf U$ as in the process tensor literature.}

\begin{figure}[hbt!]
    \centering
    \includegraphics[width=1\columnwidth]{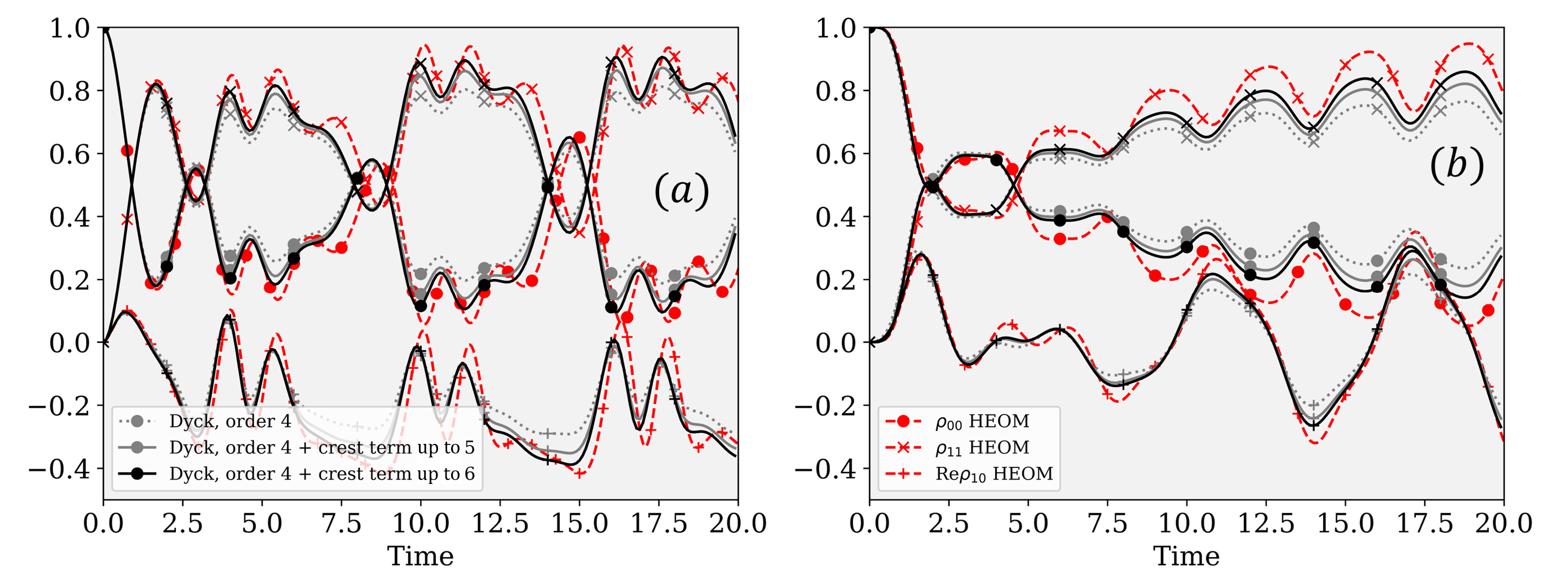}
    \caption{Dynamics of various driven systems open to a thermal bosonic environment.
    Parameters used are: $\Delta=\epsilon=1$ (other parameters are expressed relative to $\Delta=\epsilon$), $\beta=5$, $\Delta t=0.1$, $\omega_c=7.5$, and $\xi=0.1$. 
    The memory kernel is truncated at the $4$th timestep for the 
   Dyck diagrammatic method (dashed) with an additional leading term correction, the diagrams with the highest heights (smallest multiplicity), upto the sixth timestep (solid) 
   These trajectories are compared to the exact HEOM results.}
    \label{fig:driven}
\end{figure}

For problems with driven Hamiltonians, methods relying on time-translational invariance of memory kernels such as TTM become inapplicable to apply since the transfer tensor depends explicitly on time. Other schemes that treat this problem\cite{DrivenSmatpi,Pathsum} seem limited to invoking periodicity and temporal symmetries, at least in terms of numerical efficiency. Here, we devise an efficient procedure to perform the reduced system dynamics, exploiting the time-translational symmetry of the $T$s. The plan goes as follows: One computes all the $T^i$ up until a predefined truncation parameter $i=r_{\text{max}}$. Then, to obtain the kernel at time $j$, one performs a tensor contraction with the time-dependent bare system propagator tensor.

As a proof of concept, we consider various driven systems in Fig. (\ref{fig:driven}), where in panel (a): $\hat{H}_s=(\epsilon-\sin(t))\sigma_z+\sigma_x$, and in 
panel (b): $\hat{H}_s=\epsilon\sigma_z+\sin(t)\sigma_x$. In these results, we include all Dyck diagrams up to the fourth order to compute $\bm{\mathcal{K}}_0$ to $\bm{\mathcal{K}}_4$. We then approximate $\bm{\mathcal{K}}_5$ and $\bm{\mathcal{K}}_6$ by including only the diagram with the smallest multiplicity (i.e., the crest term), which is justified since the coupling here is weak and so $\tilde{I}\ll1$. We observe that while the order $4$ Dyck diagrammatic method captures the main features of the exactly computed driven dynamics (with HEOM), it is still not converged. After we include leading order correction terms up to 6 (including only the diagram with the highest height), we observe that the Dyck diagrammatic method converges to the exact values. 

In panel (a), the diagonal elements of the Hamiltonian are driven. As expected, we see a lot of interlevel population crossing here. This is to be contrasted with panel (b), where the population of $\rho_{11}$ is mostly larger than $\rho_{00}$ except for some initial time. We also observe that for the latter, the oscillations of the populations are more regular.

\section{Additional details on {\it Class 2}}\label{sec:class2}
We now start considering cases beyond our main interest, specifically beyond systems coupled to Gaussian baths where the coupling operators are all simultaneously diagonalizable.
First, as an exemplifying situation we consider a system coupled to two bosonic baths through interaction Hamiltonians whose system operators do not commute, see Ref. \citenum{palm2018quasi} for detailed analysis. The total Hamiltonian for such a setup reads
\bea
H=H_S+H_{env}^a+H_{env}^b.
\eea
We may split the total Hamiltonian as 
\bea
e^{-i\hat{H}\Delta t}\to e^{-i\hat{H}_S \Delta t} e^{-i\hat{H}_\text{env}^a \Delta t} e^{-i\hat{H}_\text{env}^b \Delta t} e^{-i\hat{H}_S \Delta t/2} + O(\Delta t^3).
\eea

We work in the diagonal $\hat{S}_b$ basis and insert resolution-of-the-identity in between to obtain
\bea
\langle b_{N}^+|\rho_\text{tot}(t)|b_{N}^-\rangle=\langle b_{N}^+| e^{-i\hat{H}_{S} \Delta t}| b_{N-1}^{+}\rangle\langle b_{N-1}^{+}| e^{-i\hat{H}_\text{env}^a \Delta t} e^{-i\hat{H}_\text{env}^b \Delta t} | b_{N-2}^{+}\rangle\langle b_{N-2}^{+}| e^{-i\hat{H}_{S} \Delta t/2}|b^+_{N-3}\rangle\cdots.
\eea
On the left of $e^{-i\hat{H}_\text{env}^b \Delta t}$ we insert resolution-of-the-identity in the $S_a$ eigenbasis. The resulting term is thus
\bea
&&\langle b_{N-1}^{+}| e^{-i\hat{H}_\text{env}^a \Delta t}\ketbra{a_{N}}{a_{N}} e^{-i\hat{H}_\text{env}^b \Delta t}| b_{N-2}^{+}\rangle\nonumber\\
&=&e^{-i\hat{H}_\text{env}^a(a_{N}^{+}) \Delta t}   e^{-i\hat{H}_\text{env}^b(b_{N-2}) \Delta t}\bra{b_{N-1}^{+}}{a_{N}}\rangle\bra{a_{N}} b_{N-2}^{+}\rangle
\eea
we therefore have
\bea
\langle b_{N}^+|\rho(t)|b_{N}^-\rangle=G_{b^\pm_{N}b^\pm_{N-1}}G_{b^\pm_{N-1}b^\pm_{N-2}}\cdots\langle b_0^+|\rho(0)|b^-_0\rangle\Tr_B[e^{-i\hat{H}_\text{env}(b_{N-1}^+)\Delta t}\cdots\rho_B \cdots e^{i\hat{H}_\text{env}(b^-_{N-1})\Delta t}]
\eea
where $G_{b^\pm_m b^\pm_{m-1}}=\langle b^+_m|e^{-i\hat{H}_{s} \Delta t/2}| b^+_{m-1}\rangle\langle b^-_{m-1}|e^{i\hat{H}_{s} \Delta t/2}| b^-_{m}\rangle$. We would then obtain that the analog to Eq. (\ref{Ut1}) is

\bea
\tilde{U}_{0,b_{0.33}^\pm b_{0.67}^\pm}&=&I_{0,b_{0.67}^{\pm}}^bI_{0,a_{0}^{\pm}}^aO_{b_{0.33}^\pm,a_0^\pm}O_{a_0^\pm,b_{0.67}^\pm}\label{Ut21}\\
\tilde{U}_{1,b_{1.67}^\pm b_{0.33}^\pm}&=&I_{0,b_{0.67}^{\pm}}^bI_{0,a_{0}^{\pm}}^aI_{0,b_{1.67}^{\pm}}^bI_{0,a_{1}^{\pm}}^aI_{1,b_{1.67}^{\pm}b_{0.67}^{\pm}}^bI_{1,a_{1}^{\pm}a_{0}^{\pm}}^aO_{b_{0.33}^\pm,a_0^\pm}O_{a_0^\pm,b_{0.67}^\pm}
O_{b_{1.33}^\pm,a_1^\pm}O_{a_1^\pm,b_{1.67}^\pm}\nonumber\\
&&G_{b^\pm_{1}b^\pm_{0.67}}G_{b^\pm_{0.33}b^\pm_{0}}\label{eq:U2class2}\\
\tilde{U}_{2,b_{2.67}^\pm b_{0.33}^\pm}&=&I_{2,b_{2.67}^{\pm}b_{0.67}^{\pm}}^bI_{2,a_{2}^{\pm}a_{0}^{\pm}}^aI_{0,b_{2.67}^{\pm}}^bI_{0,a_{2}^{\pm}}^aI_{1,b_{2.67}^{\pm}b_{1.67}^{\pm}}^bI_{1,a_{2}^{\pm}a_{1}^{\pm}}^aI_{0,b_{0.67}^{\pm}}^bI_{0,a_{0}^{\pm}}^aI_{0,b_{1.67}^{\pm}}^bI_{0,a_{1}^{\pm}}^aI_{1,b_{1.67}^{\pm}b_{0.67}^{\pm}}^bI_{1,a_{1}^{\pm}a_{0}^{\pm}}^a\nonumber\\
&&O_{b_{0.33}^\pm,a_0^\pm}O_{a_0^\pm,b_{0.67}^\pm}
O_{b_{1.33}^\pm,a_1^\pm}O_{a_1^\pm,b_{1.67}^\pm}O_{b_{2.33}^\pm,a_0^\pm}O_{a_0^\pm,b_{2.67}^\pm}\nonumber\\
&&G_{b^\pm_{1}b^\pm_{0.67}}G_{b^\pm_{0.33}b^\pm_{0}}G_{b^\pm_{2}b^\pm_{1.67}}G_{b^\pm_{1.33}b^\pm_{1}}\label{eq:U3class2}
\eea
with $O_{b_{N}^{\pm}a_{N}^\pm}=\bra{b_{N}^{\pm}}{a_{N}^\pm}\rangle$. We can similarly decompose these propagators, for example explicitly,
\bea
U_{0,b_{0}^\pm b_{1}^\pm}&=&\sum_{b_{0.67}^\pm}G_{b^\pm_{1}b^\pm_{0.67}}I_{0,b_{0.67}^{\pm}}^b\sum_{a_{0}^{\pm}}O_{a_0^\pm,b_{0.67}^\pm}I_{0,a_{0}^{\pm}}^a
\sum_{b_{0.33}^\pm}G_{b^\pm_{0.33}b^\pm_{0}}O_{b_{0.33}^\pm,a_0^\pm}
\eea
where the noninteger steps can be thought of as being auxiliary,
\bea\tilde{U}_{1,b_{1.67}^\pm b_{0.33}^\pm}&=&I_{0,b_{1.67}^{\pm}}^b\sum_{a_1^\pm}O_{a_1^\pm,b_{1.67}^\pm}I_{0,a_{1}^{\pm}}^aO_{b_{1.33}^\pm,a_1^\pm}
\sum_{b_{0.67}^\pm}F_{b^\pm_{1.33}b^\pm_{0.67}}I_{0,b_{0.67}^{\pm}}^bI_{1,b_{1.67}^{\pm}b_{0.67}^{\pm}}^b
\sum_{a_0^\pm}O_{b_{0.33}^\pm,a_0^\pm}I_{0,a_{0}^{\pm}}^aO_{a_0^\pm,b_{0.67}^\pm}I_{1,a_{1}^{\pm}a_{0}^{\pm}}^a
\nonumber\\
&=&
I_{0,b_{1.67}^{\pm}}^b\sum_{a_1^\pm}O_{a_1^\pm,b_{1.67}^\pm}I_{0,a_{1}^{\pm}}^aO_{b_{1.33}^\pm,a_1^\pm}
\sum_{b_{0.67}^\pm}F_{b^\pm_{1.33}b^\pm_{0.67}}I_{1,b_{1.67}^{\pm}b_{0.67}^{\pm}}^b
\tilde{U}_{0,b_{0.67}^\pm b_{0.33}^\pm}...
\eea
and so on. Here, it is more convenient to work with the propagators directly, and we will find
\bea
U_{2,b_{2}^\pm b_{0}^\pm}&=&\sum_{b_{1}^\pm} U_{1,b_{2}^\pm b_{1}^\pm}U_{1,b_{1}^\pm b_{0}^\pm}\nonumber\\
&&+\sum_{b_{1.67}^\pm}G_{b^\pm_{2}b^\pm_{1.67}}I_{0,b_{1.67}^{\pm}}^b\sum_{a_1^\pm}O_{a_1^\pm,b_{1.67}^\pm}I_{0,a_{1}^{\pm}}^aO_{b_{1.33}^\pm,a_1^\pm}
\sum_{b_{0.67}^\pm}F_{b^\pm_{1.33}b^\pm_{0.67}}I_{0,b_{0.67}^{\pm}}^b(I_{1,b_{1.67}^{\pm}b_{0.67}^{\pm}}^b-1)
\nonumber\\
&&\ \ \sum_{a_0^\pm}I_{0,a_{0}^{\pm}}^aO_{a_0^\pm,b_{0.67}^\pm}I_{1,a_{1}^{\pm}a_{0}^{\pm}}^a\sum_{b_{0.33}^\pm}O_{b_{0.33}^\pm,a_0^\pm}G_{b^\pm_{0.33}b^\pm_{0}}
\nonumber\\
&&+\sum_{b_{1.67}^\pm}G_{b^\pm_{2}b^\pm_{1.67}}I_{0,b_{1.67}^{\pm}}^b\sum_{a_1^\pm}O_{a_1^\pm,b_{1.67}^\pm}I_{0,a_{1}^{\pm}}^aO_{b_{1.33}^\pm,a_1^\pm}
\sum_{b_{0.67}^\pm}F_{b^\pm_{1.33}b^\pm_{0.67}}I_{0,b_{0.67}^{\pm}}^bI_{1,b_{1.67}^{\pm}b_{0.67}^{\pm}}^b
\nonumber\\
&&\ \ \sum_{a_0^\pm}I_{0,a_{0}^{\pm}}^aO_{a_0^\pm,b_{0.67}^\pm}(I_{1,a_{1}^{\pm}a_{0}^{\pm}}^a-1)\sum_{b_{0.33}^\pm}O_{b_{0.33}^\pm,a_0^\pm}G_{b^\pm_{0.33}b^\pm_{0}}
\nonumber\\
&&+\sum_{b_{1.67}^\pm}G_{b^\pm_{2}b^\pm_{1.67}}I_{0,b_{1.67}^{\pm}}^b\sum_{a_1^\pm}O_{a_1^\pm,b_{1.67}^\pm}I_{0,a_{1}^{\pm}}^aO_{b_{1.33}^\pm,a_1^\pm}
\sum_{b_{0.67}^\pm}F_{b^\pm_{1.33}b^\pm_{0.67}}I_{0,b_{0.67}^{\pm}}^b(I_{1,b_{1.67}^{\pm}b_{0.67}^{\pm}}^b-1)
\nonumber\\
&&\ \ \sum_{a_0^\pm}I_{0,a_{0}^{\pm}}^aO_{a_0^\pm,b_{0.67}^\pm}(I_{1,a_{1}^{\pm}a_{0}^{\pm}}^a-1)\sum_{b_{0.33}^\pm}O_{b_{0.33}^\pm,a_0^\pm}G_{b^\pm_{0.33}b^\pm_{0}}.
\eea
This equation will be compared with Eq. (\ref{u2minu1}), where the second, third, and fourth terms should constitute the memory kernel. The recursive, combinatorial structure of this multi-bath problem is distinct from the one-bath/multiple additive bath problems we dealt with in {\it Class 1} --detailed analysis and numerical evidence in this regard are left for future work.

\subsection{Inversion procedure for \textit{Class 2}}
For the model considered here, the inverse procedure will not follow completely the approach we outlined for \textit{Class 1}. In particular, there is no apparent diagrammatic structure underlying the expansion of $\mathbf{U}$ in terms of $\mathbf{I}$, yet \textit{in principle} it is possible to obtain $\mathbf{I}^\alpha$. To see this, we note that the knowledge of $\tilde{U}_{0,b_{0.33}^\pm b_{0.67}^\pm}$ (obtained tomographically) as well as the overlap matrices (otherwise we will obtain $\mathbf{I}^a$ ''dressed" with the overlap terms) implies $I_{0,b_{0.67}^{\pm}}^bI_{0,a_{0}^{\pm}}^a$. Here, to obtain the next order $\mathbf{I}^\alpha$ we encounter the same underdetermined problem as in the multiple bath case inversion procedure in \textit{Class 1}. So, we shall proceed here assuming the procedure is done independently-sequentially, where one of the baths' $\mathbf{I}^\alpha$ has been characterized. With this, we can thus see that we will then obtain $I_{1,b_{1.67}^{\pm}b_{0.67}^{\pm}}^bO_{a_0^\pm,b_{0.67}^\pm}$ from Eq. (\ref{eq:U2class2}) (assuming $\mathbf{I}^a$ is the one completely characterized). Similarly, we can obtain the next order from Eq. (\ref{eq:U3class2}).

\section{Additional details on {\it Class 3}}\label{sec:class3}
We exemplify problems in this class by considering the following Hamiltonian (one bath multiple interaction Hamiltonians, so the subscript $j$ is eliminated)
\bea
\hat{H}=\hat{H}_S+\hat{H}_B+\sum_\alpha\hat{S}_\alpha\otimes\hat{B}_\alpha,
\eea
where $\{\hat{B}_\alpha\}$ belong in the Hilbert space of the single bath specified by $\hat{H}_B$. For this model, let us calculate the IF, starting from Eq. (\ref{eq:A8}) while here $\mathcal{L}_{I,\alpha}$ corresponds to interaction $\alpha$ rather than bath $j$ there. The analog to Eq. (\ref{eq:A10}) hence
\bea
\rho(t)=\langle\prod_\alpha\mathcal{T}e^{\int_0^t\mathcal{L}_{I,\alpha}(\tau)d\tau}\rangle\rho(0).
\eea
If one makes an independent bath approximation, the resulting multibath influence functional would miss the correlation between different coupling terms. This correlation term is what complicates ${\it Class 3}$ compared to ${\it Class 1}$.
Now, the analog to Eq. (\ref{KGC}) is then
\bea
\langle\mathcal{T}e^{\int_0^t\sum_\alpha\mathcal{L}_{I,\alpha}(\tau)d\tau}\rangle=\sum_{n=0}^\infty\frac{1}{n!}\langle\mathcal{T}\int_0^td\tau_1\dots\int_0^td\tau_n(\sum_\alpha\mathcal{L}_{I,\alpha}(\tau_1))\dots(\sum_\alpha\mathcal{L}_{I,\alpha}(\tau_n))\rangle. \label{IFM}
\eea

Here, since the sum of independent Gaussian random variables is still Gaussian, we can apply Isserlis' theorem on Eq. (\ref{IFM}) and find that
\bea
\langle\mathcal{T}e^{\int_0^t\sum_\alpha\mathcal{L}_{I,\alpha}(\tau)d\tau}\rangle&=&\sum_{n=0}^\infty\frac{1}{(2n)!}\mathcal{T}\frac{2n!}{(2^nn!)}\Big(\int_0^td\tau_i\int_0^td\tau_k\langle (\sum_\alpha\mathcal{L}_{I,\alpha}(\tau_i))(\sum_\alpha\mathcal{L}_{I,\alpha}(\tau_k))\rangle\Big)^n\nonumber\\
&=&\sum_{n=0}^\infty\frac{1}{n!}\Big(\int_0^td\tau_i\int_0^td\tau_k\frac{\langle \mathcal{T}(\sum_\alpha\mathcal{L}_{I,\alpha}(\tau_i))(\sum_\alpha\mathcal{L}_{I,\alpha}(\tau_k))\rangle}{2}\Big)^n\nonumber\\
&=&e^{\int_0^td\tau \int_0^{\tau}d\tau'\langle \mathcal{T}(\sum_\alpha\mathcal{L}_{I,\alpha}(\tau))(\sum_\alpha\mathcal{L}_{I,\alpha}(\tau'))\rangle}\label{eq:c3}
\eea
in complete analogy to the derivation in \ref{IFG}. Here, when discretizing Eq. (\ref{eq:c3}), although one can obtain the IF, LHS of Eq. (\ref{eq:IF}), it will not factorize into pairwise separable form, the RHS of Eq. (\ref{eq:IF}). 
Thus, one needs to work with the non-pairwise separable IF. 

The inversion procedure is analogous to the one applied on {\it Class 4}, discussed in the next section. Furthermore, if one is interested to obtain the individual interaction's IF, it will not be possible to achieve this via sequential inversion in the fashion of \ref{sec:invmul}, since the correlation terms make this procedure underdetermined. Nonetheless, we can apply the technique in \ref{Appendix C F} to map this multi-interaction problem to its effective one-interaction model.

Suppose we restrict and consider a case where the system part of the interaction Hamiltonian shares a common eigenbasis and only one bath mode at a given frequency in both coupling terms. In that case, with the knowledge of the spectral density of one interaction term, we can obtain the individual interactions' spectral densities. For instance, we can obtain both spectral densities when there are two coupling terms, given access to the reduced dynamics for both couplings and one of the two couplings. This is because the correlation terms $\langle \mathcal{L}_{I,j}(\tau)\mathcal{L}_{I,k}(\tau')\rangle$ would correspond to the geometric mean spectral density, $\sqrt{J_j(\omega)}\sqrt{J_k(\omega)}$.
A detailed analysis is left for future work.

\section{Additional details on {\it Class 4}}
\label{sec:siam}
The Feynman-Vernon Path Integral formalism employed in this work was initially developed to deal with bosonic environments, due to the simple form taken by the IF in this class of models. However, recent works have extended it to fermionic systems coupled to fermionic environments\cite{ng2023real,Sonner1,Sonner2,Chen_2024,jin2010non,wmzhang}. This class of problems is best exemplified by the Single Impurity Anderson Model (SIAM), whose Hamiltonian reads
\bea
\hat{H}=  \sum_{k, \sigma} E_{k, \sigma} \hat{c}_{k, \sigma}^{\dagger} \hat{c}_{k, \sigma}+\sum_{k, \sigma}\left(V_{k} \hat{c}_{k, \sigma}^{\dagger} \hat{d}_\sigma+\text{ h.c. }\right)  +U \hat{n}_{\uparrow} \hat{n}_{\downarrow}+\sum_\sigma \varepsilon_\sigma \hat{n}_\sigma .
\eea
Note that the SIAM is related to the Kondo model through the Schrieffer–Wolff transformation. 
The main challenges to tackle this problem with the INFPI approaches as in the main text are that (1) the coupling operators are not diagonalizable nor are they simultaneously diagonalizable. (2) the fermionic nature of \textit{both} the impurity and the bath requires rethinking elementary operations (e.g., such as partial tracing) as the algebra is instead anticommutative.

First, with the same assumptions as in the main paper, taking the matrix elements of the total density matrix at time $t=N\Delta t$ (in the Grassmann coherent basis) we obtain
\bea
\bra{\overline{\boldsymbol{\xi}}_t}\rho(t) \ket{\boldsymbol{\xi}_t}=
\int \mathcal{D}[\overline{\boldsymbol{\xi}}, \boldsymbol{\xi}] \bra{\overline{\boldsymbol{\xi}}_{t_0}}\rho(t_0) \ket{\boldsymbol{\xi}_{t_0}}F[\overline{\boldsymbol{\xi}}, \boldsymbol{\xi}] \prod_\sigma \mathcal{I}_\sigma\left[\overline{\boldsymbol{\xi}}_\sigma, \boldsymbol{\xi}_\sigma\right],
\eea
There, the path integral (from $t_0$ to $t$) $\mathcal{D}[\overline{\boldsymbol{\xi}}, \boldsymbol{\xi}]$ term include the overcompleteness property of the Grassmann variables, $\mathbf{1}=\int d\mu{(\xi)}e^{-\mathbf{\xi}^*\cdot\mathbf{\xi}}\ketbra{\xi}{\xi}$. Also note that they obey Grassmann algebra $\{\xi,\hat{c}\}=0$ and the Grassmann variables themselves also anticommute. Furthermore, the term 
$F[\overline{\boldsymbol{\xi}}, \boldsymbol{\xi}]$ is the bare system propagator in the Grassmann coherent basis. In this representation, the form of the path integral is fundamentally different than that considered in our main work. Namely, due to the nilpotency of the Grassmann numbers, the variables $\xi_n$ and $\xi_n^*$ must be considered independent (and thus, compared to our main discussion these paths will be twice as long). 

Regardless, one can still devise an algebraic approach to relate the memory kernel in terms of these influence functionals even in this case because they are by construction Gaussian. In particular, the IF is calculated via taking the trace of bath degrees of freedom in the Grassmann coherent basis and is given by\cite{ng2023real,Sonner1,Sonner2,Chen_2024,jin2010non} (the integration is performed via the Grassmann gaussian integral, which is equivalent to the Hubbard-Stratonovich transformation, or via the stationary path method)
\bea
\mathcal{I}_\sigma\left[\overline{\boldsymbol{\xi}}_\sigma, \boldsymbol{\xi}_\sigma\right]=e^{-\int_{\mathcal{C}} d \tau \int_{\mathcal{C}} d \tau^{\prime} \bar{\xi}_\sigma(\tau) \Delta\left(\tau, \tau^{\prime}\right) \xi_\sigma\left(\tau^{\prime}\right)},
\eea
with 
\bea
\Delta\left(\tau, \tau^{\prime}\right)=\int \frac{d \omega}{2 \pi} J(\omega) g_{\tau, \tau^{\prime}}(\omega), \ \ J(\omega)=2 \pi \sum_k\left|V_k\right|^2 \delta\left(\omega-\epsilon_k\right), \ \ g_{\tau, \tau^{\prime}}(\omega)=\left(n_{\mathrm{F}}(\omega)-\Theta_{\mathcal{C}}\left(\tau, \tau^{\prime}\right)\right) e^{-i \omega\left(\tau-\tau^{\prime}\right)}. \label{hyb}
\eea
Discretizing this Keldysh contour requires splitting the Grassmann variables into forward and backward branches. In the limit of a small timestep, the hybridization function takes a simple matrix form, see Refs. \citenum{Sonner1,Sonner2,Chen_2024}. Yet, to our knowledge, the discretized IF will not take a pairwise form precisely because of the Grassmann algebra. As a result, we only show the formal relationship between the IFs and the memory kernel, which is algebraic in structure. Future work will focus on finding a geometric and diagrammatic structure between the two for the SIAM.

In this model, we will find that (setting $\overline{\boldsymbol{\xi}}_j,\boldsymbol{\xi}_j\to\boldsymbol{\Xi}_j$ for brevity)
\bea
\tilde{\mathbf{U}}_0[{\boldsymbol{\Xi}_1}]&=&\mathcal I[\boldsymbol{\Xi}_1],\nonumber\\
\tilde{\mathbf{U}}_1[{\boldsymbol{\Xi}_3\boldsymbol{\Xi}_1}]&=&{F}[{\boldsymbol{\Xi}_3} \boldsymbol{\Xi}_1]\mathcal 
 I[\boldsymbol{\Xi}_3\boldsymbol{\Xi}_1],\nonumber\\
\tilde{\mathbf{U}}_2[{\boldsymbol{\Xi}_5\boldsymbol{\Xi}_1}]&=&
\sum_{\boldsymbol{\Xi}_3}
{F}[{\boldsymbol{\Xi}_5} \boldsymbol{\Xi}_5]
{F}[{\boldsymbol{\Xi}_3} \boldsymbol{\Xi}_1]\mathcal 
 I[\boldsymbol{\Xi}_5\boldsymbol{\Xi}_3\boldsymbol{\Xi}_1], \label{SIAMU}
\eea
and thus formally, for example,
\bea
\mathcal{K}_1[{\boldsymbol{\Xi}_4} \boldsymbol{\Xi}_0]&=&\frac{1}{\Delta t^2}(\sum_{\boldsymbol{\Xi}_3,\boldsymbol{\Xi}_1}{G}[{\boldsymbol{\Xi}_4} \boldsymbol{\Xi}_3]{F}[{\boldsymbol{\Xi}_3} \boldsymbol{\Xi}_1]\mathcal 
 I[\boldsymbol{\Xi}_3\boldsymbol{\Xi}_1]{G}[{\boldsymbol{\Xi}_1} \boldsymbol{\Xi}_0]\nonumber\\
 &&-\sum_{\boldsymbol{\Xi}_3,\boldsymbol{\Xi}_2,\boldsymbol{\Xi}_1}{G}[{\boldsymbol{\Xi}_4} \boldsymbol{\Xi}_3]\mathcal I[{\boldsymbol{\Xi}_3}]
 {G}[{\boldsymbol{\Xi}_3} \boldsymbol{\Xi}_2]
 {G}[{\boldsymbol{\Xi}_2} \boldsymbol{\Xi}_1]
 \mathcal I[{\boldsymbol{\Xi}_1}]
 {G}[{\boldsymbol{\Xi}_1} \boldsymbol{\Xi}_0])
 , \label{SIAMK}
\eea
where $\mathcal 
 I[\boldsymbol{\Xi}_n\dots\boldsymbol{\Xi}_1]$ defines the IF along the Grassmann path (note further that we also abbreviate $\boldsymbol{\xi}_n=(\boldsymbol{\xi}_{\uparrow,n},\boldsymbol{\xi}_{\downarrow,n})$) from both spin sectors.

\subsection{Inversion procedure for the SIAM}
Similarly, for the inversion procedure of problems in \textit{Class 2}, the inverse procedure will not follow completely the approach we outlined earlier in this section. We sketch the necessary modifications here. First, we assume the density matrix has been tomographically obtained in the coherent Grassmann basis. Furthermore, despite the lack of the Dyck path structure in Eqs. \ref{SIAMU}, for the first few steps, we shall also take that we have obtained $\mathcal{K}$ in the sense of Eq. \ref{SIAMK}. Hence, by rearranging the series of equations for $\mathcal{K}_N$, we can obtain $\mathcal 
I[\boldsymbol{\Xi}_1],\mathcal 
I[\boldsymbol{\Xi}_3\boldsymbol{\Xi}_1],\mathcal 
I[\boldsymbol{\Xi}_5\boldsymbol{\Xi}_3\boldsymbol{\Xi}_1]\dots$. Thus, recalling that $\overline{\boldsymbol{\xi}}_j,\boldsymbol{\xi}_j\to\boldsymbol{\Xi}_j$ and $\boldsymbol{\xi}_n=({\xi}_{\uparrow,n},{\xi}_{\downarrow,n})$, by this point we have\cite{Chen_2024} (with a slight abuse of notation) 
\bea
\mathcal{I}_{\sigma,0}&=&e^{-\sum_{\zeta,\zeta'=\pm}\bar{\xi}_{\sigma,1}^\zeta\Delta_{1,1}^{\zeta,\zeta'} \xi_{\sigma,1}^{\zeta'}}=1-\sum_{\zeta,\zeta'=\pm}\bar{\xi}_{\sigma,1}^\zeta\Delta_{1,1}^{\zeta,\zeta'} \xi_{\sigma,1}^{\zeta'},\label{ISIAM1}\nonumber\\
\mathcal{I}_{\sigma,1}&=&e^{-\sum_{j,k}^2\sum_{\zeta,\zeta'=\pm}\bar{\xi}_{\sigma,j}^\zeta\Delta_{j,k}^{\zeta,\zeta'} \xi_{\sigma,k}^{\zeta'}}=1-\sum_{j,k}^2\sum_{\zeta,\zeta'=\pm}\bar{\xi}_{\sigma,j}^\zeta\Delta_{j,k}^{\zeta,\zeta'} \xi_{\sigma,k}^{\zeta'},\label{ISIAM2}\nonumber\\
&\vdots&
\eea
where $\zeta,\zeta'=\pm$ denotes the forward-backward Keldysh branch after discretization. Now, to recollect, since we have full knowledge of the set $\mathcal 
I[\boldsymbol{\Xi}_1],\mathcal 
I[\boldsymbol{\Xi}_3\boldsymbol{\Xi}_1],\mathcal 
I[\boldsymbol{\Xi}_5\boldsymbol{\Xi}_3\boldsymbol{\Xi}_1]\dots$ (up to a truncation order, LHS of Eqs. \ref{ISIAM1}) and following, and the Grassmann numbers $\overline{\boldsymbol{\xi}}_{\sigma,j}^\zeta,\boldsymbol{\xi}_{\sigma,j}^{\zeta'}$, we, therefore, know the hybridization matrices $\Delta_{jk}^{\zeta,\zeta'}$. This object is analogous to $\eta_{\Delta k}$ in e.g., the spin-boson case, with the difference that $\Delta_{jk}^{\zeta,\zeta'}$ takes into account the fermionic nature of the bath and the system (this comes from the $\Theta_{\mathcal{C}}$ term in Eq. (\ref{hyb})). Thus, one can proceed in the same fashion, continuing from Eq. (\ref{etaJ}) to obtain the fermionic spectral density from Eq. (\ref{hyb}) by performing the appropriate Fourier Transform.

\section{HEOM Calculations \label{sec:HEOM}}
All hierarchical equations of motion (HEOM) \cite{doi:10.1143/JPSJ.58.101,doi:10.1063/5.0011599} calculations presented in this work made use of the free pole HEOM variant,\cite{FPHEOM} which makes use of the adaptive Antoulas--Anderson (AAA) algorithm\cite{doi:10.1137/16M1106122} for constructing a simple rational function approximation to the bath noise spectrum,
\begin{equation}
    S(\omega) = \frac{1}{2}J(\omega) \left[\coth\left(\frac{\beta\omega}{2}\right) + 1\right]
\end{equation}
which, in turn, gives rise to a sum of exponential functions representation of the bath correlation function
\begin{equation}
    C(t) \approx \sum_{j=1}^K \alpha_k e^{ - \nu_k t}, 
\end{equation}
with $\mathrm{Re}(\nu_k) \geq 0$.
The HEOM approach makes use of the Feynman-Vernon influence functional discussed in \ref{sec:pi}, to obtain a description of the exact dynamics of an open quantum system in terms of an infinite hierarchy of auxiliary density operators (ADOs),  $\hat{\rho}_{\boldsymbol{m}, \boldsymbol{n}}(t)$, that encode the system-bath correlations.  These ADOs are indexed by two sets of integers $\boldsymbol{m} = (m_{0}, m_{1}, \dots, m_{K})$ and $\boldsymbol{n} = (n_{0}, n_{1}, \dots, n_{K})$, each of length $K$.\cite{FPHEOM}
This hierarchy is truncated for practical calculations such that no element of $\boldsymbol{m}$ or $\boldsymbol{n}$ is greater than some integer $L$.

\begin{figure}[hbt!]
    \centering
    \includegraphics[width=0.75\columnwidth]{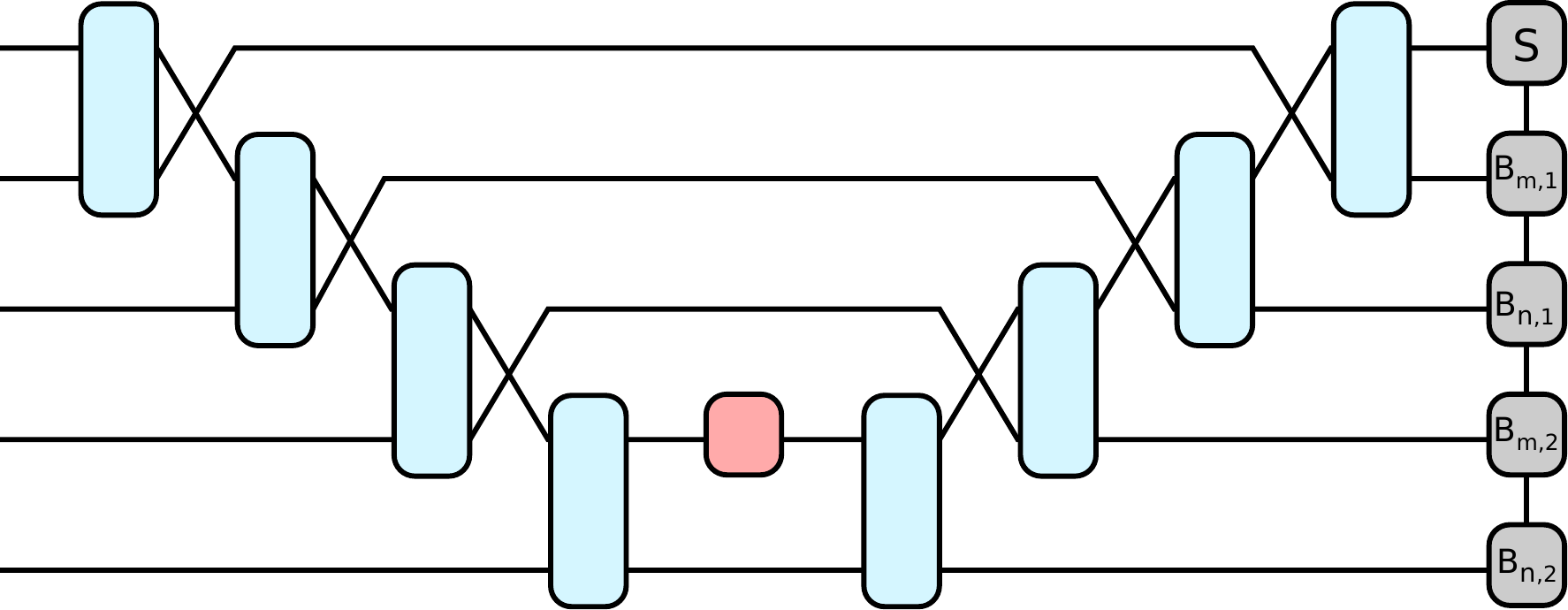}
    \caption{An illustration of a single time step of the swap-based two-site TEBD algorithm used to integrate the FP-HEOMs.  Crossing of lines indicates a swap operation applied to the MPS state.  The grey rectangles represent the MPS representation of the hierarchy of ADOs, $\hat{\boldsymbol{\rho}}$.  The light blue rectangles correspond to the short-time bath propagators, $\mathcal{B}_{m,k}$, $\mathcal{B}_{n,k}$, and red rectangles correspond to the short-time system propagator term, $\mathcal{U}_S(t)$.
    \label{fig:swapTEBD}}
\end{figure}

Within the FP-HEOM method, this hierarchy of ADOs evolves according to the equations of motion\cite{FPHEOM}
\begin{equation}
\begin{split}
    \frac{\partial}{\partial t}\hat{\rho}_{\boldsymbol{m}, \boldsymbol{n}}(t) =& - \left(i\mathcal{L}_s(t) + \sum_{k=1}^K (\nu_k m_k + \nu_k^* n_k) \right)\hat{\rho}_{\boldsymbol{m}, \boldsymbol{n}}(t) - i\sum_{k=1}^N \left(\sqrt{\frac{m_k}{\|\alpha_k\|}} \alpha_k \hat{S} \hat{\rho}_{\boldsymbol{m}_k^-, \boldsymbol{n}}(t)  - \sqrt{\frac{n_k}{\|\alpha_k\|}} \alpha_k^*\hat{\rho}_{\boldsymbol{m}, \boldsymbol{n}_k^-}(t)\hat{S}\right) \\ & - i\sum_{k=1}^N \left(\sqrt{(m_k +1)\|\alpha_k\|} \left[\hat{S}, \hat{\rho}_{\boldsymbol{m}_k^+, \boldsymbol{n}}(t)\right]  + \sqrt{(n_k +1)\|\alpha_k\|} \left[\hat{S}, \hat{\rho}_{\boldsymbol{m}, \boldsymbol{n}_k^+}(t)\right]\right), 
\end{split}\label{eq:HEOM}
\end{equation}
where $\boldsymbol{m}_k^\pm$($\boldsymbol{n}_k^\pm$) corresponds to the set of indices $\boldsymbol{m}$($\boldsymbol{n}$) but with the $k$-th element incremented (+) or decremented (-) by one.
These equations have the general form
\begin{equation}    \begin{split}
    \frac{\partial}{\partial t}\hat{\boldsymbol{\rho}}(t) = & -i\mathcal{L}_s(t)\hat{\boldsymbol{\rho}}(t) + \sum_{k=1}^N \left[\mathcal{L}_{m,k} + \mathcal{L}_{n,k}\right]\hat{\boldsymbol{\rho}}(t) \\
     =& \mathcal{M}(t) \hat{\boldsymbol{\rho}}(t) ,
    \end{split}
\end{equation}
where $\mathcal{L}_{m,k}$ is an operator that acts on system and the $k$-th mode of the hierarchy of ADOs.

For the ohmic and subohmic spectral densities with exponential cutoffs considered in this work, the total number of exponentials in the sum of exponential representation of the bath correlation function leads to a set of auxiliary density operators that are too large to represent exactly.  Following references \onlinecite{10.1063/1.5026753, FPHEOM, Mangaud2023} we use a matrix product state (MPS) representation of the ADOs. Evolution of the HEOMs is performed through the use of a two-site time-evolving block decimation (TEBD) algorithm that makes use of a symmetric Trotter splitting of the short-time HEOM propagator,
\begin{equation}
\begin{split}
    \mathcal{U}_{\mathrm{heom}}(t, t+\Delta t) &= \mathcal{T}\exp\left(\int_t^{t+\Delta t}\mathcal{M}(\tau) \mathrm{d}\tau\right) \\ 
    &\approx \left(\prod_{k=1}^K \exp\left[ \mathcal{L}_{m, k} \frac{\Delta t}{2}\right] \exp\left[ \mathcal{L}_{n, k} \frac{\Delta t}{2}\right]\right) \exp\left[-i \mathcal{L}_S(t+\frac{\Delta t}{2}) \frac{\Delta t}{2}\right]\left(\prod_{k=K}^1 \exp\left[ \mathcal{L}_{n, k} \frac{\Delta t}{2}\right] \exp\left[ \mathcal{L}_{m, k} \frac{\Delta t}{2}\right]\right) \\ 
    &= \left(\prod_{k=1}^K  \mathcal{B}_{m, k}\mathcal{B}_{n, k} \right) \mathcal{U}_s(t)\left(\prod_{k=K}^1 \mathcal{B}_{n, k} \mathcal{B}_{m, k} \right)
\end{split}\label{eq:HEOM_prop}
\end{equation}
In contrast to the single-site time-dependent variational principle-based schemes that have previously been used for the time evolution of tensor network-based approximations to the HEOMs,\cite{10.1063/1.5026753,lindoythesis, FPHEOM, Mangaud2023,10.1063/5.0153870} the use of a two-site TEBD scheme naturally allows for adaptive control of the MPS bond dimension throughout a simulation.

The interactions present in the HEOMs given in Eq. \ref{eq:HEOM} have a star topology; terms acting on two system modes all act on the system degree of freedom as well as one of the modes of the hierarchy.  The application of the propagator in Eq. \ref{eq:HEOM_prop}  would require the application of two-body operators acting on modes that are not nearest-neighbor in the MPS, giving rise to an algorithm that requires $O(K^2)$ two-site updates at each time step.  To avoid this overhead, we employ a strategy where at the application of each two-body term, we swap the position of the system and bath site being acted upon, ensuring that at each stage the next two-body operator to apply is a local operation on the MPS.\cite{PhysRevX.7.031013,Bauernfeindthesis}  This update scheme is illustrated schematically in Fig. \ref{fig:swapTEBD}, and requires $O(K)$ two-site updates at each time step. This scheme has the advantage that we perform a TEBD update at each stage between the system site and a single bath site. Consequentially, when the system Liouville space dimension is much smaller than the hierarchy depth, the cost of two-site updates will be dramatically reduced compared to the naive approach.

\begin{figure}[hbt!]
    \centering
    \includegraphics[width=1\columnwidth]{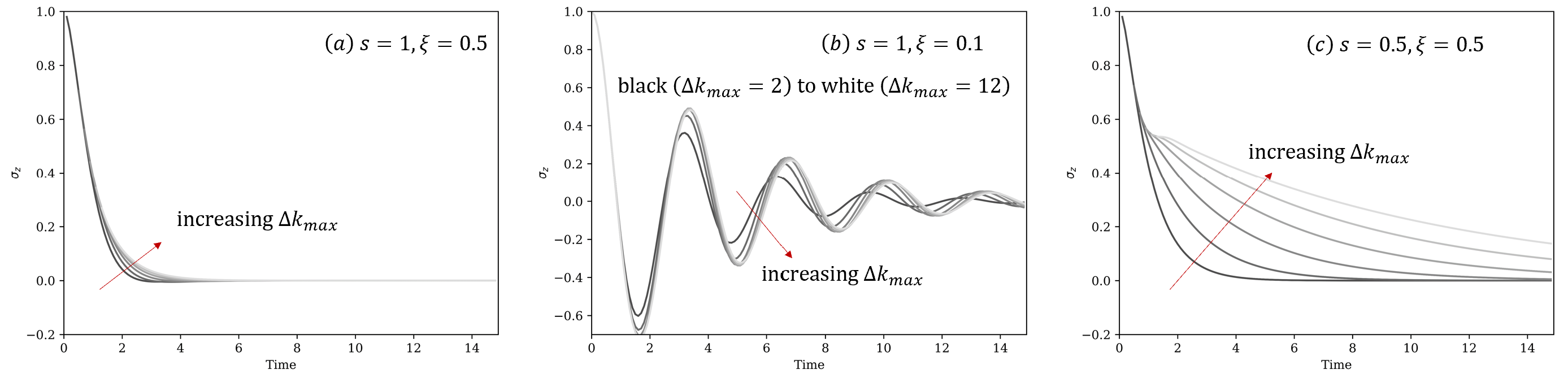}
    \caption{Magnetization ($\langle \sigma_z(t)\rangle$) dynamics predicted using i-QuAPI method with increasing $\Delta k_{\text{max}}$ (from black to white colors, $\Delta k_{\text{max}}=2$ to $\Delta k_{\text{max}}=12$.). Parameters used are: $\Delta=1$, $\epsilon=0$, $\beta=5$, $\Delta t=0.1$, $\omega_c=7.5$, and $\xi=0.1$ and $s=1$ (panel (a)),
    $\xi=0.5$ and $s=1$ (panel (b)), or $\xi=0.5$ and $s=0.5$ (panel (c))}
    \label{fig:convergence}
\end{figure}

\section{Additional Simulations}
\label{sec: Appendix F}
For Figures (\ref{Numerics1}) and (\ref{Numerics2}) panels (a) and (b), the exact results from which our calculations compare come from converged HEOM results. We cross-validate these results using the i-QuAPI method, which agrees with HEOM, see Fig. (\ref{fig:convergence}).

\begin{figure}[hbt!]
    \centering
    \includegraphics[width=0.33\columnwidth]{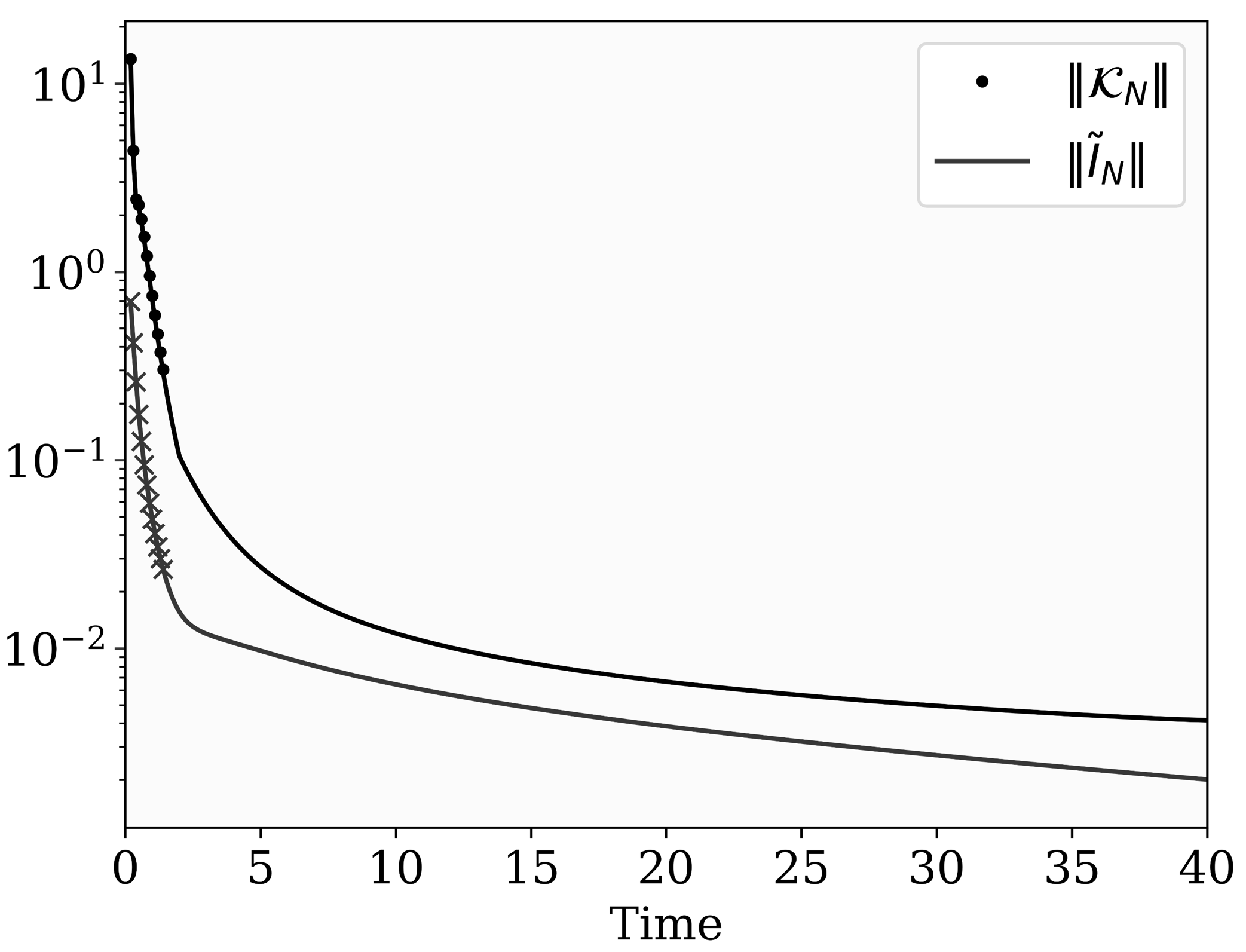}
    \caption{Same as \cref{Numerics1} with a longer time.}
    \label{fig:long}
\end{figure}

Fig. (\ref{Numerics1}) panel (c) suggests that the decay of $\norm{\tilde{\mathbf{{I}}}_N}$ (and hence $\norm{\bm{\mathcal{K}}_N}$) is extremely slow. One question is if they eventually decay to zero at all. In Fig. (\ref{fig:long}) we present $\norm{\tilde{\mathbf{{I}}}_N}$ and $\norm{\bm{\mathcal{K}}_N}$ for longer times than in the main text. We observe that neither has vanished and both are still decaying even until $t = 40$.

In Fig. (\ref{fig:eta}), we demonstrate the perfect agreement between inversely obtained $\eta$ to the analytical $\eta$. This is the intermediate step to extract the spectral density of the environment; see \ref{Appendix C F}.

Furthermore, we show how we can inversely obtain $J(\omega)$ for highly structured environments in Fig. \ref{fig:structured}: in panel (a) $J_a(\omega)=\frac{\gamma \omega \omega_0^2}{(\omega^2-\omega_0^2)^2+\gamma^2\omega^2}+\frac{\gamma \omega \omega_0^2}{(\omega^2-4\omega_0^2)^2+\gamma^2\omega^2}$, for panel (b) $J_b(\omega)=\frac{\gamma}{(\omega-\omega_0)^2+\gamma^2}+\frac{\gamma}{(\omega-3\omega_0)^2+\gamma^2}$, and for panel (c) $J_c(\omega)=\frac{\gamma \omega \omega_0^2}{(\omega^2-\omega_0^2)^2+\gamma^2\omega^2}+\frac{\gamma}{(\omega-3\omega_0)^2+\gamma^2}$. Specifically these are spectral densities that are the additions of Brownian-Brownian, Lorentz-Lorentz, and Brownian-Lorentz spectral densities with different peak frequencies, respectively. It is significant that even for highly structured spectral densities such as these, the extraction procedure succeeds, although one would need to go to high orders in the Dyck diagrams.

Lastly, we look at how the truncation in $\bm{\mathcal{K}}$ affects $\mathbf I$ and vice versa. We construct approximate $\bm{\mathcal{K}}$ with truncated $\tilde{\mathbf{{I}}}$ (i.e., akin to i-QUAPI). Similarly, we extract effective $\mathbf{{I}}$ with truncated $\bm{\mathcal{K}}$ (i.e., akin to GQME). We present these results in Fig. (\ref{fig:approximate}). Here, we study the three different regimes considered in the main text. In panels (a1), (b1), and (c1), we show the decay of $\norm{\bm{\mathcal{K}}_N}$ of Dyck order $13$, if we truncate (set $\tilde{\mathbf{{I}}}_{k,ij}=0$ for $k>k_{\text{max}}$) at $k_{\text{max}}=5$, $k_{\text{max}}=9$, and $k_{\text{max}}=13$ respectively. Here, one observes the error of premature truncation compounds, at e.g., $t=1.2$ the error of $\norm{\bm{\mathcal{K}}_N}$ when truncating at $k_{\text{max}}=5$ is significantly larger than when truncating at $k_{\text{max}}=9$. On the other hand, it appears that the effective $\tilde{\mathbf{{I}}}$ extracted from this truncated $ \bm{\mathcal{K}}$ is a poor approximate of the actual $\tilde{\mathbf{{I}}}$ (although this makes sense since we impose a hard truncation, $\tilde{\mathbf{{I}}}_{k,ij}=0$ for $k>k_{\text{max}}$). This is shown in panels (a2), (b2), and (c2); of Dyck order $13$, at $k_{\text{max}}=5$, $k_{\text{max}}=9$, and $k_{\text{max}}=13$, respectively.

\begin{figure}[hbt!]
    \centering
    \includegraphics[width=0.9\columnwidth]{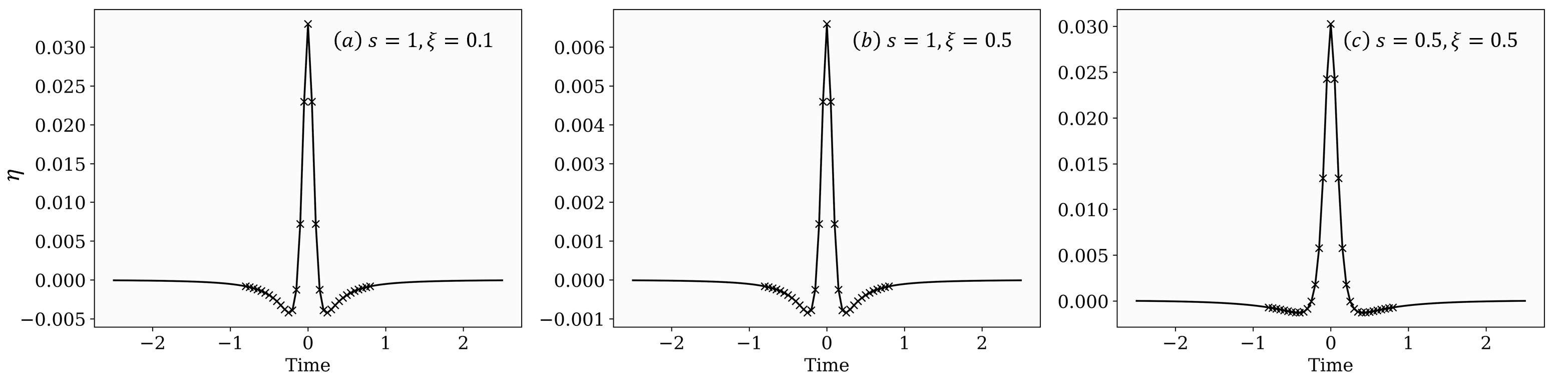}
    \caption{The coefficients $\eta_{\Delta k}$ inversely calculated (solid line), which perfectly matches the a priori results (crosses). Parameters are identical to \cref{Numerics2} in the main text.}
    \label{fig:eta}
\end{figure}
\begin{figure}[hbt!]
    \centering
    \includegraphics[width=1\columnwidth]{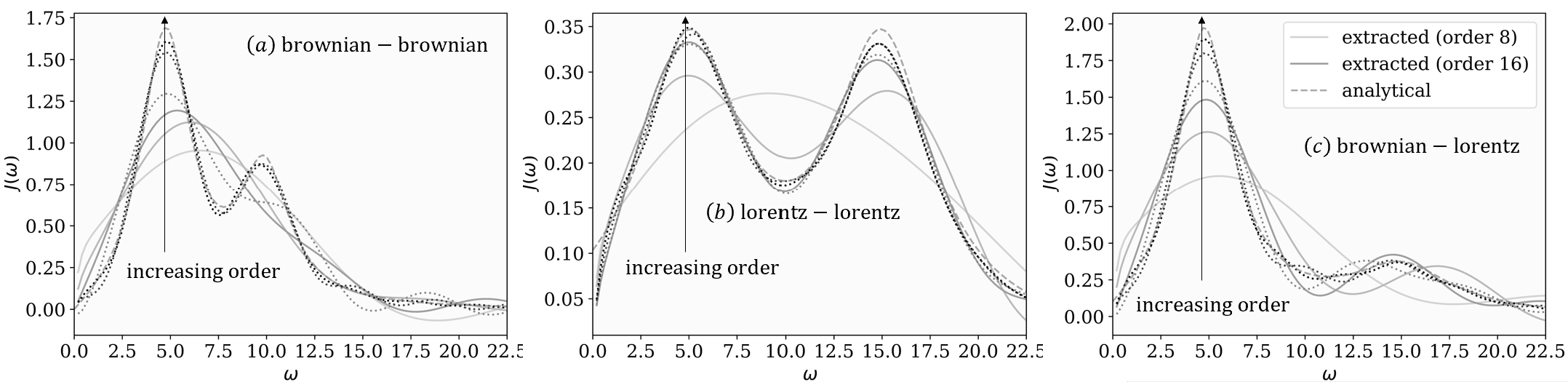}
    \caption{
    Panels (a), (b), and (c): Bath spectral densities extracted through the Dyck diagrammatic method with increasing truncation orders (from white to black colors, $8,12,16,20,30,$ and $40$ in sequence) compared to exact spectral densities (dashed), see \ref{Appendix C F}. Note that for orders $20, 30,$ and $40$ (dotted), we directly proceed from $\mathbf{I}\to\eta\to J(\omega)$ since the procedure $\bm{\mathcal{K}}\to\mathbf{I}$ for these high orders are currently infeasible (see paragraph below Eq. (\ref{I2jn})). In all cases, our approach obtains the structured spectral density. For panel (a) the exact spectral density $J_a(\omega)=\frac{\gamma \omega \omega_0^2}{(\omega^2-\omega_0^2)^2+\gamma^2\omega^2}+\frac{\gamma \omega \omega_0^2}{(\omega^2-4\omega_0^2)^2+\gamma^2\omega^2}$, for panel (b) $J_b(\omega)=\frac{\gamma}{(\omega-\omega_0)^2+\gamma^2}+\frac{\gamma}{(\omega-3\omega_0)^2+\gamma^2}$, and for panel (c) $J_c(\omega)=\frac{\gamma \omega \omega_0^2}{(\omega^2-\omega_0^2)^2+\gamma^2\omega^2}+\frac{\gamma}{(\omega-3\omega_0)^2+\gamma^2}$.
    Parameters used are: $\Delta=1$ (other parameters are expressed relative to $\Delta$), $\epsilon=0$, $\beta=5$, $\Delta t=0.05$, $\omega_c=7.5$, $\xi=2$, $\gamma=\frac{\xi \pi}{2}$, and $\omega_0=5$.
    }
    \label{fig:structured}
\end{figure}

\begin{figure}[hbt!]
    \centering
    \includegraphics[width=1\columnwidth]{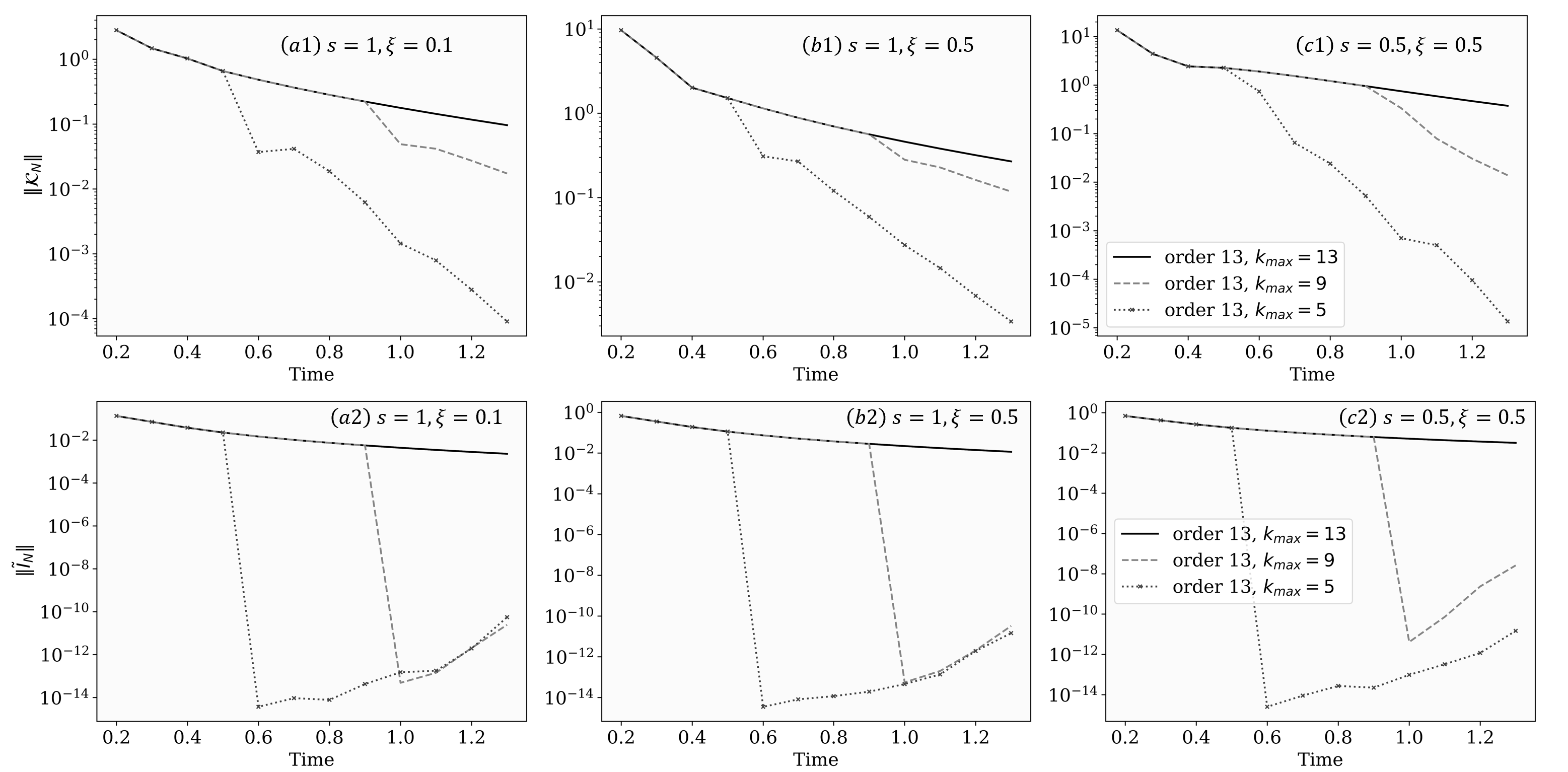}
    \caption{Panels (a1), (b1), and (c1): The operator norm of the approximate $\bm{\mathcal{K}}_N$ when we truncate (set $\tilde{I}_{k,ij}=0$ for $k>k_{\text{max}}$) at $k_{\text{max}}=5$, $k_{\text{max}}=9$, and $k_{\text{max}}=13$, respectively. The solid line is the exact $\bm{\mathcal K}_N$. Panels (a2), (b2), and (c2): The operator norm of $\tilde{\mathbf I}_N$ obtained from approximate $\bm{\mathcal{K}}_N$ shown in panels (a1), (b1), and (c1). The solid line indicates the exact $\tilde{\mathbf I}_N$.
     Parameters used are: $\Delta=1$ (other parameters are expressed relative to $\Delta$), $\epsilon=0$, $\beta=5$, $\Delta t=0.1$, $\omega_c=7.5$, and $\xi=0.1$ and $s=1$ (panel (a)), 
    $\xi=0.5$ and $s=1$ (panel (b)), or $\xi=0.5$ and $s=0.5$ (panel (c))
    }
    \label{fig:approximate}
\end{figure}
\end{widetext}
\end{document}